\documentclass[
10pt,
twocolumn,
secnumarabic,
nofootinbib,
amsmath,amssymb,
aps,
pra,
longbibliography
]{revtex4-2}

\usepackage{amsmath}
\usepackage{centernot}
\usepackage{amsmath, amssymb}
\usepackage{amsthm}
\usepackage{mathrsfs}
\usepackage{braket,mleftright}
\usepackage{dsfont}
\usepackage{bbm}
\usepackage{float}
\usepackage{comment}
\usepackage{graphicx}% Include figure files
\usepackage{dcolumn}% Align table columns on a decimal point
\usepackage{bm}% bold math
\usepackage[colorlinks=true, urlcolor=blue, linkcolor=blue, citecolor=blue]{hyperref}% add hypertext capabilities
\usepackage{xcolor}
\usepackage{url}
\usepackage{lipsum}
\usepackage{mathtools}
\usepackage{easybmat}
\usepackage{multirow}
\usepackage{diagbox}
\usepackage{booktabs}

\newcommand{\RNum}[1]{\expandafter{\romannumeral #1\relax}}

\begin{document}

\title{Duality breaking, mobility edges, and the connection between topological Aubry-Andr{\'e} and quantum Hall insulators
in atomic wires with fermions} 

\author{Bar Alluf}
\affiliation{School of Physics, Georgia Institute of Technology, Atlanta, 30332, USA}

\author{C. A. R. {S\'a} de Melo}
\affiliation{School of Physics, Georgia Institute of Technology, Atlanta, 30332, USA}
\date{\today}

\begin{abstract}
It is well known that the Aubry-Andr{\'e} model lacks mobility edges due to its energy-independent self-duality but may exhibit edge states. 
When duality is broken, we show that mobility regions arise and non-trivial topological phases emerge.
By varying the degree of duality breaking, we identify mobility regions and establish a connection between Aubry-Andr{\'e} atomic wires with fermions and quantum Hall systems for a family of Hamiltonians that depends on the relative phase of laser fields, viewed as a synthetic dimension.
Depending on the filling factor and the degree of duality breaking, we find three classes of non-trivial phases: conventional topological insulator, conventional topological Aubry-Andr{\'e} insulator, and unconventional (hybrid) topological Aubry-Andr{\'e} insulator. Finally, we discuss appropriate Chern numbers that illustrate the classification of topological phases of localized fermions in atomic wires.

\end{abstract}

\maketitle

\section{Introduction}
\label{sec:introduction}

The concept of mobility edge pertains to an energy boundary that distinguishes between localized and extended states.
The existence of mobility edges (MEs) holds immense significance in comprehending intricate localization phenomena observed in diverse physical systems, such as Anderson insulators \cite{anderson-1958}, photonic materials \cite{righini-1997, genack-2000, maret-2006, mordechai-2007, silberberg-2008, bouyer-2012, christodoulides-2013, szameit-2018, scheffold-2020},
kicked rotor systems \cite{delande-2015},
flatband networks \cite{flach-2014},
and ultracold atoms with speckles \cite{bouyer-2006, aspect-2007, aspect-2008, DeMarco-2011, hutchinson-2020}.

In one dimension, an infinitesimal amount of random disorder is sufficient to cause localization and MEs are not present \cite{mott-1987}.
Mobility edges are also absent in the Aubry-Andr{\'e} (AA) model \cite{andre-1980}, describing localization in one-dimensional (1D) quasiperiodic potentials, due to a global duality that directly maps extended states to localized states and vice versa.
There are some generalizations of the AA model that produce MEs: the addition of mean-field interactions \cite{bloch-2018, sarma-2015, gadway-2021, suotang-2022, datta-2022, yu-2022}, modified quasiperiodicities \cite{liu-2020}, beyond nearest-neighbor hopping \cite{sarma-2009, sarma-2010}, and noninteracting systems with spin-orbit coupling (SOC) but without Rabi fields \cite{tobe-2008, zhang-2013, konotop-2022}.
Furthermore, MEs may also be found in 1D speckle-disordered systems, such as ${}^{87}\rm{Rb}$, where it was shown that SOC and Rabi fields facilitate transport and hinder localization \cite{spielman-2020}.
Motivated by the experimental observation of MEs in Aubry-Andr{\'e} atomic wires including mean-field interactions with bosons \cite{gadway-2021, suotang-2022}, we propose a mechanism for breaking the global duality of Aubry-Andr{\'e} systems, by introducing SOC and Rabi fields, that produces not only MEs but also topological order in atomic wires with fermions.
Our global duality-breaking mechanism can also be applied to study many-body localization of ultracold fermions in optical lattices \cite{aidelsburger-2021a, aidelsburger-2021b}.
Using the current platform of atomic wires with fermions \cite{aidelsburger-2021a, aidelsburger-2021b} and adding duality-breaking fields, we investigate the phase diagrams for different filling factors, characterize the emergent mobility regions, and identify topological states.

Depending on the sequence of MEs, we find three types of
topological phases with midgap edge states. The first phase is
a conventional topological insulator with a bulk gap separating two conducting bands. The second one is a conventional
topological AA insulator with two insulating bands separated
by a bulk gap. The third phase is an unconventional (hybrid)
topological AA insulator with a bulk gap separating insulating
and conducting bands. The last phase is a direct consequence
of the existence of MEs. To characterize these topological
phases, we introduce a fictitious dimension (the phase difference between weak and strong lasers) in AA atomic wires.
We also compute the topological invariants (Chern numbers)
of each phase by mapping our problem into a 2D system
describing generalized quantum Hall insulators.
The remainder of the paper is organized as follows.
In \ref{sec3.2} we describe the Hamiltonian used in our analysis.
In \ref{sec3.3} we discuss the localization properties of our Hamiltonian by analyzing the inverse participation ratio.
In \ref{sec3.4} we perform a scaling analysis of the inverse participation ratio to be used for the determination of phase diagrams.
In \ref{sec3.5} we investigate various phase diagrams showing localized and extended regions.
In \ref{sec3.6} we describe the connection between quantum Hall systems with spin-orbit and Rabi couplings and obtain Chern numbers to identify
topological phases.
In \ref{sec3.7} we compare our work with
previous efforts.
In \ref{sec3.8} we summarize and present our conclusions.

\section{Lattice Hamiltonian}
\label{sec3.2}

We start with the duality-broken AA Hamiltonian
\begin{equation}
    \mathcal H_\varphi = -\sum_{\braket{nm}ss'} J_{nm}^{ss'} c_{ns}^\dagger c_{ms'} + \sum_{nss'} \Gamma_{nn}^{ss'} c_{ns}^{\dagger} c_{ns'} 
    \label{eq3.1}
\end{equation}
where $c_{ns}^\dagger$ and $c_{ns}$ describe fermions at lattice site $n$ with spin $s$ and $\braket{nm}$ represents nearest neighbors.
The first matrix is
\begin{equation}
    \boldsymbol{J}_{nm} = J_{nm}[\cos(k_T \delta x_{nm})\mathbf{I} + i\sin(k_T\delta x_{nm}\boldsymbol{\sigma}_z)],
    \label{eq3.2}
\end{equation}
describing nearest-neighbor hopping, where $k_T$ is the spindependent momentum transfer associated with SOC and $\delta x_{nm} = x_n - x_m$ are displacements, where $x_n = na$, with $a$ the optical lattice spacing determined by the wavelength $\lambda_S$ of a strong laser \cite{gadway-2021, suotang-2022}.
Here $\mathbf I$ is the identity and $\boldsymbol{\sigma}_i$ is the Pauli matrix for $i \in {x,y,z}$.
The second matrix is
\begin{equation}
    \boldsymbol{\Gamma}_{nn} = \Delta \cos(2\pi \beta n - \varphi) \mathbf{I} - h_x \boldsymbol{\sigma}_x,
    \label{eq3.3}
\end{equation}
where $\Delta$ is a modulation created by a weak laser beam 
\cite{gadway-2021, suotang-2022} of wavelength $\lambda_W$, $\varphi = 2(\phi_S - \phi_W)$ is twice the phase difference between the strong $\phi_S$ and
weak $\phi_W$ lasers, $\beta = \lambda_S / \lambda_W$ is the ratio of wavelengths, and $h_x$ plays the role of the Rabi field.
For definiteness, we set $\beta = 532/738$, compatible with experiments with ${}^{40}\rm {K}$ \cite{bloch-2015}.

In addition, using the spin-gauge transformation (SGT) $\mathbf{c}_n = e^{ik_T x_n\boldsymbol{\sigma}_z} \tilde{\mathbf{c}}_n$ with $\mathbf{c}_n = (c_{n\uparrow}, c_{n\downarrow})^T$, where $T$ means transposition, we transfer $\exp(ik_T \delta x_{nm})$ from $\boldsymbol{J}_{nm}$ to $\boldsymbol{\Gamma}_{nn}$, leading to
\begin{equation}
    \tilde{\mathcal{H}}_\varphi = -\sum_{\braket{nm}ss'} \tilde{\boldsymbol{J}}_{nm}^{ss'} \tilde{c}_{ns}^\dagger \tilde{c}_{ms'} + \sum_{nss'} \tilde{\Gamma}_{nn}^{ss'} \tilde{c}_{ns}^\dagger \tilde{c}_{ns'}.
    \label{eq3.4}
\end{equation}
Here $\tilde{J}^{ss'}_{nm}$ are the matrix elements of
\begin{equation}
    \tilde{\boldsymbol{J}}_{nm} = J_{nm} \mathbf I,
    \label{eq3.5}
\end{equation}
where $\tilde{\boldsymbol{J}}_{nm}$ does not contain spin-dependent phases.
The local spin-diagonal elements $\tilde{\Gamma}^{ss}_{nn} = \Gamma^{ss}_{nn}$ remain invariant; however, the spin-off-diagonal elements become $\tilde{\Gamma}_{nn}^{\uparrow\downarrow} = \Gamma_{nn}^{\uparrow\downarrow} e^{-2i k_T x_n} = -h_x e^{-2i k_T x_n}$ and $\tilde{\Gamma}_{nn}^{\downarrow\uparrow} = \Gamma_{nn}^{\downarrow\uparrow} e^{2i k_T x_n} = -h_x e^{2i k_T x_n}$.
Using $x_n = na$, we write the matrix elements $\tilde{\Gamma}_{nn}^{ss'}$ in matrix form as
\begin{equation}
    \tilde{\boldsymbol{\Gamma}}_{nn} = \Delta \cos(2\pi \beta n - \varphi) \mathbf I - h_+(n) \boldsymbol{\sigma}_+ -h_-(n) \boldsymbol{\sigma}_-,
    \label{eq3.6}
\end{equation}
where $\boldsymbol{\sigma}_{\pm} = (\boldsymbol{\sigma}_x \pm i\boldsymbol{\sigma}_y)/2$ are spin raising $(+)$ and lowering $(-)$ matrices and
\begin{equation}
    h_{\pm}(n) = h_x e^{\mp i2k_T an}
    \label{eq3.7}
\end{equation}
are the helical Rabi fields with spatial variation controlled by the SOC parameter $k_T a$.
The matrix in \ref{eq3.6} can also be written in terms of the Pauli matrices $\sigma_x$ and $\sigma_y$ as
\begin{equation}
    \tilde{\boldsymbol{\Gamma}}_{nn} = \Delta \cos(2\pi\beta n-\varphi) \mathbf I - h_x(n)\boldsymbol{\sigma}_x - h_y(n)\boldsymbol{\sigma}_y,
    \label{eq3.8}
\end{equation}
where the local Rabi field components are
\begin{equation}
    h_x(n) = h_x \cos(2k_T an)
    \label{eq3.9}
\end{equation}
in the $x$ direction in spin space and 
\begin{equation}
    h_y(n) = h_x \sin(2k_T an)
    \label{eq3.10}
\end{equation}
in the $y$ direction in spin space.
Note that the SGT makes explicit in Eqs.~(\ref{eq3.4})-(\ref{eq3.10}) that the Hamiltonian $\tilde{\mathcal H}_\varphi$ is $\pi$ periodic in $k_T a$ and that for $h_x = 0$ the system’s energy is independent
of the spin-orbit parameter $k_T a$, due to spin-gauge symmetry.
Throughout the text, we define the filling factor as $\nu =
N_{st}/N$, where $N_{st}$ is the number of states and $N$ is the number of sites.
Using this definition, $\nu$ ranges from 0 to 2, that is $0 \le \nu \le 2$.
Thus, $\nu = 1/2$ corresponds to quarter filling, $\nu=1$ labels half filling, and $\nu=2$ describes full filling.
Since we are interested in the localization properties of our
Hamiltonian as energy $E/J$ or filling factor $\nu$ change, we
investigate next the inverse participation ratio (IPR) for changing $E/J$ and $\nu$ as a function of the Hamiltonian parameters $\Delta/J$, $k_T a$, and $h_x/J$, where we take the hopping $J$ as our energy unit.

\section{Localization Properties}
\label{sec3.3}

\begin{figure}
    \centering
    \includegraphics[width=\linewidth]{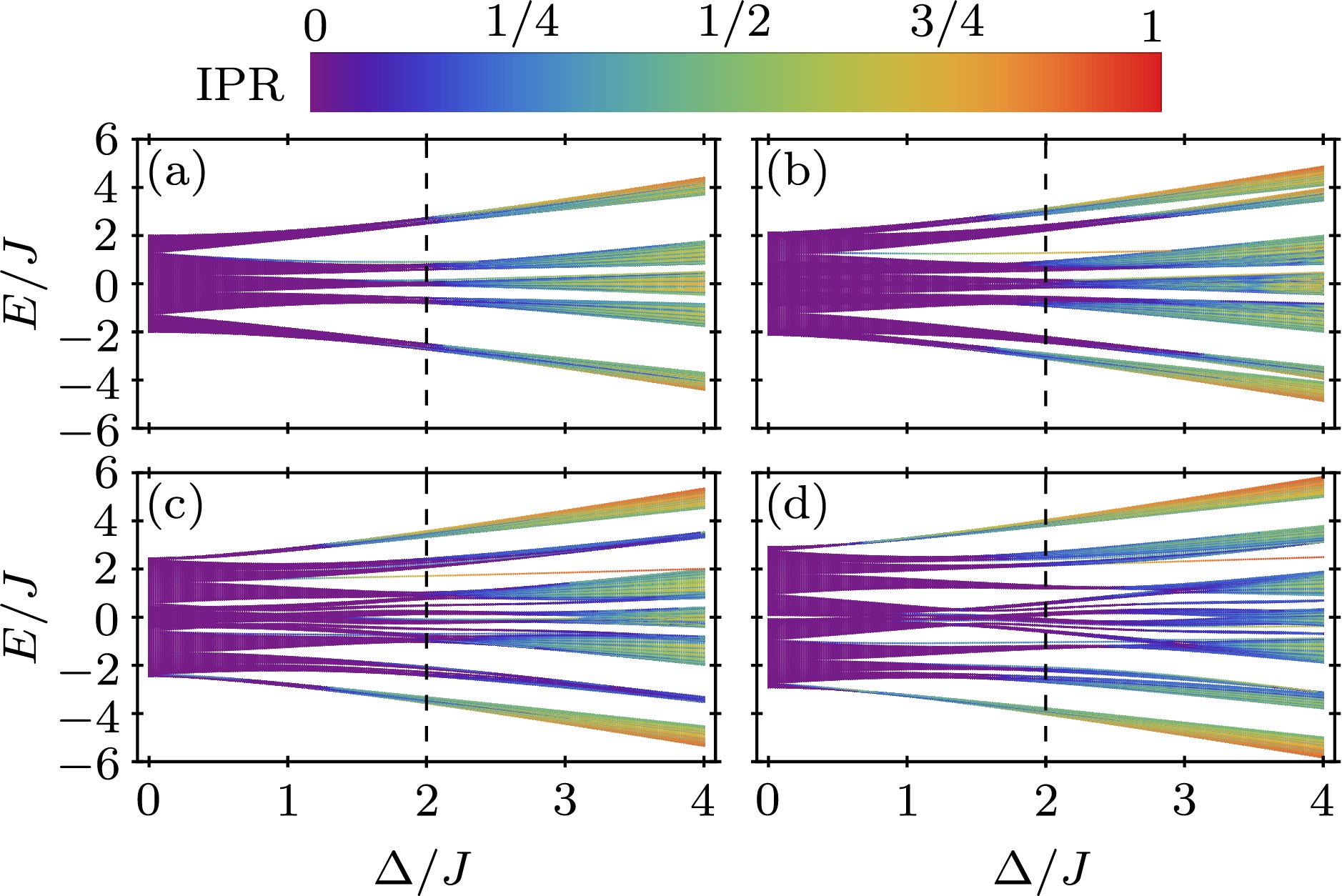}
    \caption{Plots of $E/J$ versus $\Delta/J$ showing mobility edges for SOC and Rabi coupling with $\beta = 532 / 738$, $N=501$, $k_T a = \pi/4$, and $\varphi = 0$.
    The black dashed lines represent $(\Delta/J)_c^{\rm{AA}}=2$.
    The parameters are (a) $h_x/J = 0$, (b) $h_x/J = 0.5$, (c) $h_x/J = 1$, (d) $h_x/J = 1.5$. 
    The continuous IPR color scheme varies from violet indicating extended states to red indicating localized states.
    In (a) there is a sharp transition at $(\Delta/J)_c^{\rm{AA}}=2$ and no mobility edges, while in (b)-(d) the changing Rabi fields lead to mobility edges both below and above $(\Delta/J)_c^{\rm{AA}}=2$.}
    \label{fig1}
\end{figure}

We use exact diagonalization to compute the inverse participation ratio
\begin{equation}
    \rm{IPR} = \sum_i \chi_i^2 = \sum_i \big( \vert \psi_{i\uparrow} \vert^2 + \vert \psi_{i\downarrow} \vert^2 \big)^2
    \label{eq3.11}
\end{equation}
where $\chi_i = \sum_s \vert \psi_{is}\vert^2$ and the normalization condition $\sum_{is} \vert\psi_{is}\vert^2=1$.
The IPR is used to classify energy eigenstates as localized or extended.
For a fully extended state $\chi_i = 1/N$, where $N$ is the number of sites, the IPR is equal to $1/N$, going to zero in the thermodynamic limit of $N\to \infty$.
For a fully localized state at site $j$, we obtain $\chi_i = \delta_{ij}$, and the IPR equals 1.
For finite-size systems, the IPR range is $1/N \le \rm{IPR} \le 1$.
When $(N\to\infty)$ 
the IPR range is 
$0 \le \rm{IPR} \le 1$. \ref{sec3.4} we discuss further the scaling behavior of the IPR with system size $L=(N-1)a$ and its relation to the localization length $\xi$.
Duality plays an important role in the localization properties of bichromatic lattices.
The conventional AA model is self-dual and does not exhibit mobility edges or regions \cite{andre-1980}, while the generalized Aubry-Andr{\'e} (GAA) system has an energy-dependent self-duality that produces a mobility edge \cite{bloch-2018, sarma-2015, gadway-2021, suotang-2022, datta-2022}.
In contrast to the last two examples, the presence of SOC and Rabi fields breaks the AA model's self-duality, revealing the much richer behavior seen in Fig.~\ref{fig1}.

A heat map of the IPR is shown in Fig.~\ref{fig1} for the energy
$E/J$ versus the quasiperiodic modulation $\Delta/J$ plane at a fixed value of the SOC parameter $k_T a = \pi/4$ and four values of the Rabi field $h_x/J$.
The physics revealed in Fig.~\ref{fig1} is that, in one-dimensional
optical lattices, spin-orbit coupling, and Rabi fields globally
break duality but only affect particle-hole symmetry weakly
through the boundary conditions.
As a result, mobility regions rather than mobility edges arise. Since the part of the
Hamiltonian that contains spin-orbit coupling and Rabi fields
is particle-hole symmetric in the bulk, this implies that when it
induces a mobility edge for particles, it also induces a mobility
edge for holes at fixed values of $\Delta / J$.
The full Hamiltonians $\mathcal H_\varphi$ in Eq.~[\ref{eq3.1}] and $\tilde{\mathcal H}_\varphi$ in Eq.~[\ref{eq3.4}] are not particle-hole symmetric but are nearly so due to the sparsity of edge states.
This near particle-hole symmetry produces mobility regions rather than simply mobility edges as found in situations where particle-hole symmetry is strongly broken.
For example, the potential of the GAA model
\begin{equation*}
    V_n = \frac{\Delta \cos(2\pi n \beta + \phi)}{1-\alpha \cos(2\pi n \beta + \phi)}
\end{equation*}
discussed in Refs. \cite{bloch-2018, sarma-2015, gadway-2021, suotang-2022, datta-2022}, where $n$ is the site index, breaks
particle-hole symmetry but preserves an energy-dependent
duality that produces an analytical mobility edge
\begin{equation*}
    \frac{E}{J} = \frac{1}{\alpha}\left(2-\frac{\Delta}{J}\right),
\end{equation*}
assuming $\Delta > 0$ and $J > 0$.
When $\alpha < 0$, the mobility edge arises below half filling $\nu < 1 \ (E < 0)$ for $\Delta / J < 2$ and above half filling $\nu > 1 \ (E>0)$ for $\Delta / J > 2$. 
When $\alpha > 0$, the converse is true, that is, the mobility edge arises above half filling $\nu > 1 \ (E>0)$ for $\Delta / J < 2$ and below half filling $\nu < 1 \ (E<0)$ for $\Delta / J > 2$.
In the case of spin-orbit coupling and Rabi fields, we do not obtain an analytical expression for the boundaries of the mobility regions, because self-duality is explicitly broken, but we emphasize that the near particle-hole symmetry of the full Hamiltonian $\mathcal H_\varphi$ or $\tilde{\mathcal H}_\varphi$ produces mobility regions that are almost particle-hole symmetric in contrast to the GAA model.

\begin{figure}
    \centering
    \includegraphics[width=\linewidth]{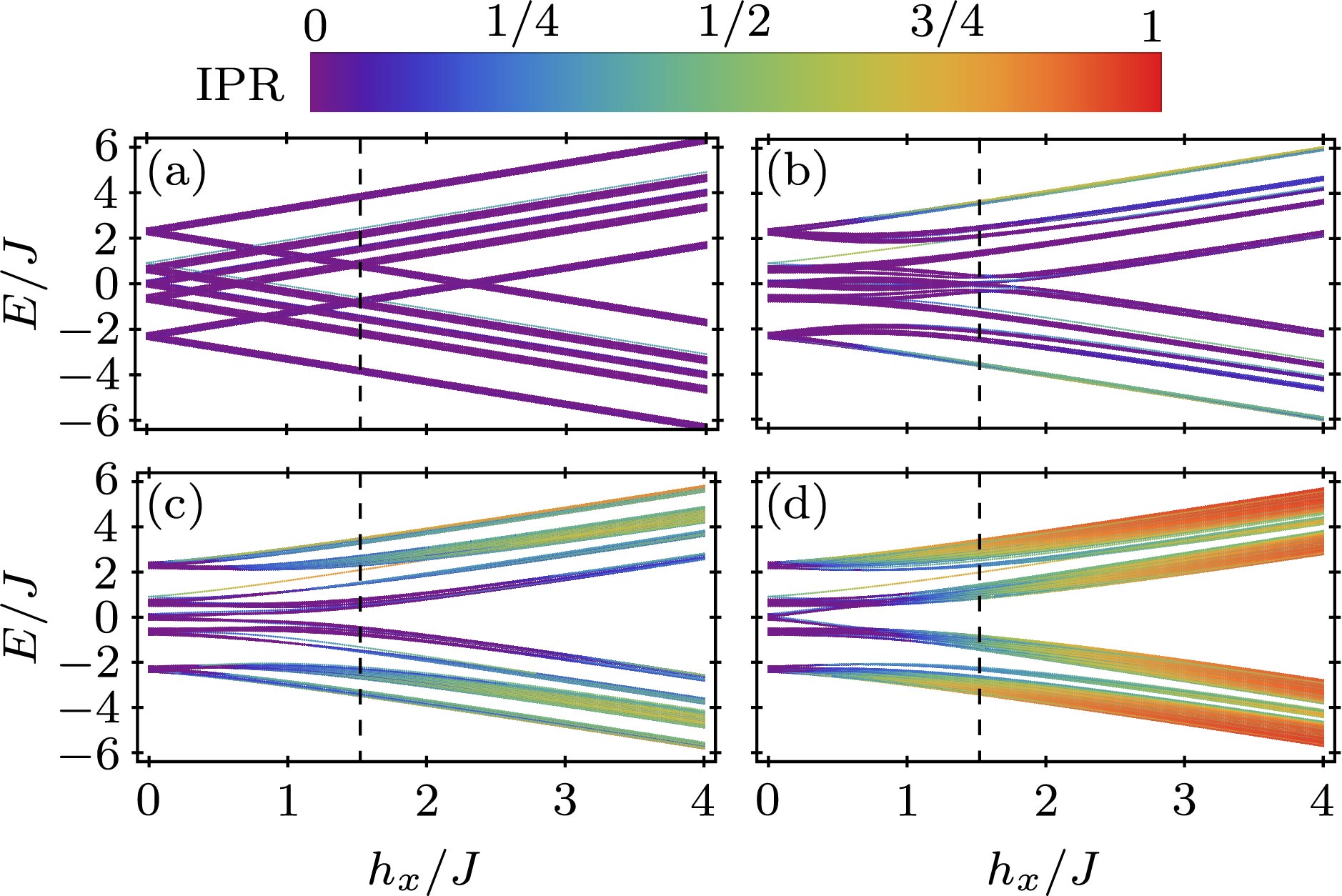}
    \caption{Plots of $E/J$ versus $h_x/J$ with $\beta = 532 / 738$, $N=501$, $\Delta/J = 1.5$, and $\varphi = 0$.
    The SOC parameters are (a) $k_T a = 0$, (b) $k_T a = \pi/4$, (c) $k_T a = 3\pi/8$, and (d) $k_T a = \pi/2$.
    The continuous color IPR varies from extended violet to localized red.
    When $k_T a \ne 0$, the low- and high-energy states localize faster than intermediate ones.
    The black dashed lines represent $h_x/J = 1.5$.
    }
    \label{fig2}
\end{figure}

In Fig.~\ref{fig1} we show a density plot of the IPR in the energy $E/J$ versus $\Delta/J$ plane for $k_T a = \pi/4$ and various $h_x$.
We set $\beta = 532/738$, the number of sites $N = 501$, and the
phase $\psi = 0$.
The other parameters are $h_x/J = 0$ [Fig.~\ref{fig1}(a)], $h_x/J = 0.5$ [Fig.~\ref{fig1}(b)], $h_x/J = 1$ [Fig.~\ref{fig1}(c)], and $h_x/J = 1.5$ [Fig.~\ref{fig1}(d)].
The black dashed lines indicate the critical ratio $(\Delta/J)_c^{\rm{AA}}=2$ beyond which all the states are localized in the conventional AA model.
The violet (red) lines represent extended (localized) states, most of which are in the bulk, but a few localized states in the white gaps are located at the edges.
Notice, in all panels, the nearly particle-hole symmetric spectra about $E/J = 0$, correspond to half-filling $(\nu = 1)$ and the existence of mobility regions (violet) where the eigenstates are extended.
Notice also the occasional edge state that migrates from one energy bundle to another as $\Delta/J$ changes.
In Fig.~\ref{fig1}(a), where $h_x/J = 0$, the spin-gauge symmetry leads to the duality-preserving AA model, which undergoes a phase transition at $\Delta / J = 2$ for any energy $E/J$.
No mobility edge is present: All states with $\Delta/J <2$ are extended and all states with $\Delta/J > 2$ are localized.
In Fig.~\ref{fig1}(b), where $h_x / J = 0.5$, spin-gauge symmetry breaking also leads to self-duality breaking.
Some states with high and low energies (filling factors) localize below $\Delta/J = 2$, while some states with intermediate energies (filling factors) are still delocalized above $\Delta/J = 2$.
Similar behaviors occur in Fig.~\ref{fig1}(c) for $h_x/J = 1$ and Fig.~\ref{fig1}(d) for $h_x/J = 1.5$.
Fig.~\ref{fig1}(b)-(d), for fixed $\Delta/J$, show mobility regions, with leading and trailing mobility edges, due to the duality-breaking introduced by SOC and Rabi fields and its near preservation of particle-hole symmetry.
In contrast, only a single mobility edge exists in the GAA model, as observed in bosonic AA wires [22,23], where an energy-dependent self-duality is preserved when mean-field interactions are considered, and particle-hole symmetry is strongly broken.
To provide further insight into the localization properties of
the Aubry-Andr{\'e} model in the presence of spin-orbit coupling
and Rabi fields, we set the value of $\Delta/J = 1.5$, which is
below the Aubry-Andr{\'e} localization threshold $(\Delta/J)_c^{\rm{AA}}=2$ and investigate the effects of the Rabi field $h_x/J$ on the IPR of states with different energies $E/J$ for a few values of the SOC parameter $k_T a$.
Thus, in Fig.~\ref{fig2} we show density plots of the IPR in the energy $E/J$ versus $h_x/J$ plane.
Fully extended states appear in violet $(\rm{IPR \to 0})$ and fully localized states appear in red $(\rm{IPR \to 1})$.
Here $\beta = 532/738$, $\Delta/J = 1.5$, system size $N = 501$, and phase $\varphi = 0$ for different values of SOC: $k_T a = 0$ [Fig.~\ref{fig2}(a)], $k_T a = \pi/4$ [Fig.~\ref{fig2}(b)],
$k_T a = 3\pi/8$ [Fig.~\ref{fig2}(c)], and $k_T a = \pi/2$ [Fig.~\ref{fig2}(d)].
The white background areas describe insulating regions.
The occasional midgap lines, connecting upper and lower bands, are edge states.
We will discuss the importance of these states \ref{sec3.6}.
The physical interpretation of the panels in Fig.~\ref{fig2} is as follows.
When the spin-orbit parameter $k_T a = 0$, self-duality is preserved as there are two copies of the local quasiperiodic
potential with the Rabi field being uniform in the $x$ direction in spin space, as seen in Eqs.~(\ref{eq3.8})-(\ref{eq3.10}): For $k_T a = 0$, the local Rabi fields are uniform $h_x (n) = h_x$ and $h_y (n) = 0$, the local energy matrix is $\tilde{\Gamma}_{nn} = \Delta \cos(2\pi\beta n - \varphi) \mathbf I - h_x \boldsymbol{\sigma}_x$, and the nearest-neighbor hopping matrix is $\tilde{\boldsymbol{J}}_{n,n+1} = J \mathbf I$ from Eq.~[\ref{eq3.5}].

The first (second) copy of the quasiperiodic potential is shifted downward (upward) in energy by $-h_x (+h_x)$, but no mobility regions arise due to the preservation of self-duality for each copy.
This means that for $\Delta/J < 2 (\Delta/J>2)$, all states must be extended (localized).
This physics is shown in Fig.~\ref{fig2}(a), where all states are extended since $\Delta/J = 1.5$ is below the Aubry-Andr{\'e} localization threshold $(\Delta/J)_c^{\rm{AA}}$.
However, when both $k_T a$ and $h_x$ are nonzero, self-duality is broken and mobility regions, instead of a simple mobility edge, arise due to the near particle-hole symmetric Hamiltonian $\mathcal H_\varphi (\tilde{\mathcal H_\varphi})$.
Figures~\ref{fig2}(b)-(d) show the effect of the Rabi field $h_x$ in determining the IPR and the boundaries of mobility regions for fixed nonzero values of $k_T a$.
For any nonzero values of $k_T a$ the eigenvaluesof the local energy matrix $\tilde{\Gamma}_{nn}$ are
\begin{equation}
    \varepsilon_n = \Delta \cos(2\pi n \beta - \varphi) \pm h_x,
    \label{eq3.12}
\end{equation}
and in the basis that diagonalizes $\tilde{\Gamma}_{nn}$ the nearest-beighbor hoping matrix is
\begin{equation}
    \tilde{\boldsymbol{J}}_{n,n+1} = J\big[ \cos(k_T a)\mathbf I - i\sin(k_Ta)\boldsymbol{\sigma}_x \big]
    \label{eq3.13}
\end{equation}
These two equations show that it becomes increasingly more difficult to hop from site $n$ to site $n+1$ as $h_x$ gets larger in the local energies $\varepsilon_n$.
This means that along a particular eigenvalue line $E/J$, the IPR tends to increase with $h_x/J$ as localization is facilitated.
For fixed $k_T a\ne 0$, the states with low particle (hole) energy, that is, with low (high) eigenvalues $E/J$ or low (high) filling factors $\nu$, tend to localize first with $h_x/J$ increases, while the states with high particle (hole) energy that is, with low (high) eigenvalues $E/J$ or low (high) filling factors $\nu$, tend to localize first when $h_x/J$ increases, while the states with high particle (hole) energy, that is, with eigenvalues $E/J \approx 0$ or filling factors $\nu \approx 1$, tend to localize last when $h_x/J$ increases.

Fig.~\ref{fig2} also shows that mobility (violet) regions shrink as $k_T$ a increases from 0 to $\pi/2$, that states corresponding to energies (filling factors) about $E/J \approx 0 (\nu \approx 1)$ are more robust to localization, and that states with high and low energies (filling factors $\nu \approx 0$ or 2) localize more easily.
The physical reason for this effect is that the local helical Rabi field in Eqs.~(\ref{eq3.9}) and (\ref{eq3.10}) varies from uniform, with $h_x (n) = h_x$ and $h_y (n) = 0$ for any site $n$ (when $k_T a = 0$)
to staggered, with $h_\pm(n) = h_x e^{\pm i \pi n} = h_x (-1)^n$, that is, $h_x (n) = h_x$ and $h_y = 0$, for $n$ even,
while
$h_x(n) = -h_x$ and $h_y = 0$
for $n$ odd (when $k_Ta=\pi/2$).
This implies that, in the staggered case, it is more
difficult for particles (holes) to hop from site to site by means of a spin-independent nearest-neighbor hopping
$\tilde{\boldsymbol{J}}_{n,n+1} = J \mathbf I$, defined in Eq.~[\ref{eq3.5}].

\begin{figure}
    \centering
    \includegraphics[width=\linewidth]{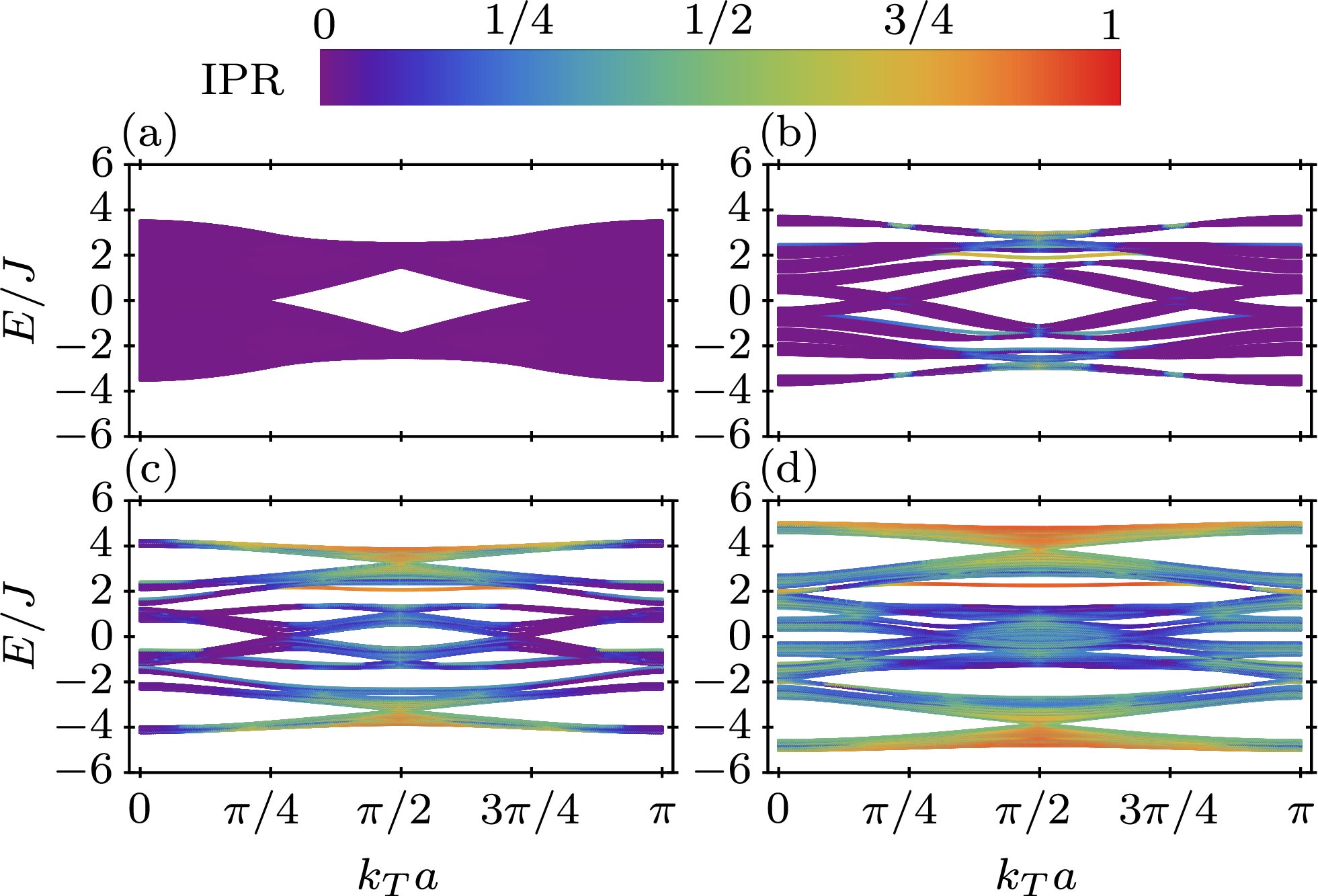}
    \caption{Plots of $E/J$ versus $k_T a$ with $\beta = 532 / 738$, $N=501$, $h_x/J = 1.5$, and $\varphi = 0$.
    The paramaters are (a) $\Delta/J = 0$, (b) $\Delta/J = 1$, (c) $\Delta/J = 2$, and (d) $\Delta/J = 3$.
    The continuous color IPR varies from extended violet to localized red.
    When $h_x a \ne 0$, eigenstates tend to have higher IPR at $k_T a = \pi/2$.
}
    \label{fig3}
\end{figure}

To illuminate further the effects of the SOC parameter $k_T a$,
Fig.~\ref{fig3} shows a density plot of the IPR in the plane $E/J$ versus $k_T a$ for fixed $\beta = 532/738$, $h_x/J = 1.5$, $N = 501$, and $\varphi = 0$, but a few values of $\Delta/J$.
In all panels of Fig.~\ref{fig3}, $E/J$ and the IPR are $\pi$ periodic in $k_T a$, a property that is made explicit by the spin-gauge transformation discussed in Eqs.~(\ref{eq3.4})-(\ref{eq3.10}),
where it is shown that the Hamiltonian $\tilde{\mathcal H}_\varphi$ is periodic operator in $2 k_T a$, that is, $\tilde{\mathcal H}_\varphi$ is $\pi$ periodic with respect to $k_T a$.
In
Fig.~\ref{fig3}(a), $\Delta/J = 0$ and all states are extended for any $k_T a$.
This is physically due to the spin-gauge symmetry that reduces $\tilde{\mathcal H}_\varphi$ to the standard Aubry-Andr{\'e} model. Since $\Delta/J$ is
below the critical value for localization, that is, $\Delta/J = 1.5 < (\Delta /J)_c^{\rm{AA}} = 2$, all states are extended.
In Fig.~\ref{fig3}(b)–(d) the parameters are $\Delta/J = 1, 2,$ and 3, respectively, with $h_x/J = 1.5$.
Notice that localization occurs first at $k_T a = \pi /2 (\rm{mod}\pi)$, where the local Rabi field is staggered, making it difficult for the eigenfunctions with low eigenvalues (low-energy particles) and high eigenvalues (lowenergy holes) to be extended.
As a result, larger values of the IPR for lower eigenvalues (lower-energy particles) and higher eigenvalues (lower-energy holes) arise.
Due to the staggering of the local Rabi field, the $h_x/J = 1.5$ is sufficient to localize the eigenfunctions for lower and higher eigenvalues at $k_T a = \pi /2$.
In contrast, when $k_T a = 0$, the local Rabi fields are uniform for $h_x/J \ne 0$, reducing our model to two spin copies
of the Aubry-Andr{\'e} model, where the eigenstates are always
extended for $\Delta / J < 2$.
For fixed $\Delta / J$, the helical nature of the local Rabi field with components $h_x(n)$ and $h_y(n)$ and its dependence on $k_T$ a is largely responsible for the increase of the IPR from extended (violet) towards localized (red), at a given energy state, as $k_T a$ changes from 0 to $\pi/2$.
For instance, this general tendency is clearly seen in Fig.~\ref{fig3}(c) and (d).
Furthermore, at fixed $h_x/J = 1.5$, a small $k_T a \ne 0$ may
help either localize or delocalize eigenstates depending on the
value of $\Delta / J$ and filling factor $\nu$.
Fig.~\ref{fig3}(b)–(d) show clear mobility regions, instead of simple mobility edges, in the $E/J$ versus $k_T a$ plane, due to the near particle-hole symmetry of the Hamiltonian $\tilde{\mathcal H}_\varphi$.
Sparse edge states also appear in white energy gaps of Fig.~\ref{fig3}(b)–(d).
The importance of these edge states are discussed in \ref{sec3.6}.
We discussed above localization properties and mobility regions caused by duality-breaking spin-orbit coupling and Rabi fields and nearly preserving particle-hole symmetry of the Hamiltonian $\tilde{\mathcal H}_\varphi$ in Eq.~[\ref{eq3.4}].
Next we perform a scaling analysis of our results and obtain localization-delocalization phase diagrams for various externally controllable pairs of parameters.

\section{Scaling Analysis}
\label{sec3.4}

To better understand the localization properties of the
eigenstates of our Hamiltonian $\tilde{\mathcal H}_\varphi$, we perform an extensive study of the IPR as a function of the number of sites $N$.
First, we show that the IPR values calculated at $N = 501$, with system size $L = (N-1)a$, have converged to the desired relative precision.
Second, we analyze the scaling of the IPR with the system size $L$.
Third, we show that the IPR shows the appropriate scaling behavior with respect to the localization length $\xi$ and system size $L$.

\subsection{IPR convergence}
\begin{figure}
    \centering
    \includegraphics[width=\linewidth]{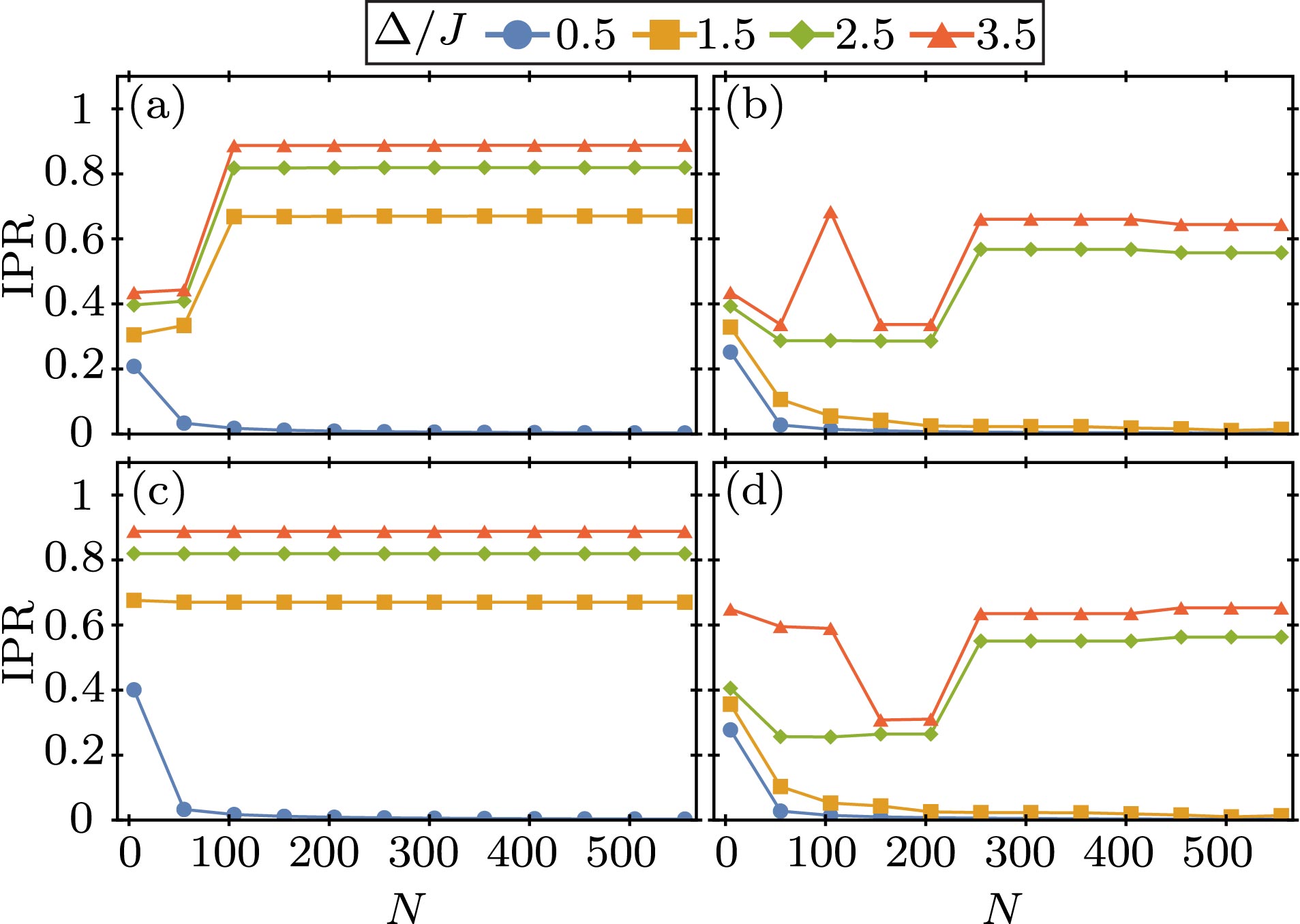}
    \caption{Plots of the IPR versus the system size $N$ for $\beta = 532/738$, $k_T a = \pi/4$, $h_x/J = 3$, and $\varphi = 0$.
    In all panels, we use $\Delta/J = \{0.5, 1.5, 2.5, 3.5\}$, where (a) $\nu \approx 0$ (ground state), (b) $\nu \approx 1/2$ (quarter-filled state), (c) $\nu \approx 2$ (highest-energy state), and (d) $\nu \approx 3/2$ (3/4-filled state).}
    \label{fig4}
\end{figure}

\begin{figure}
    \centering
    \includegraphics[width=\linewidth]{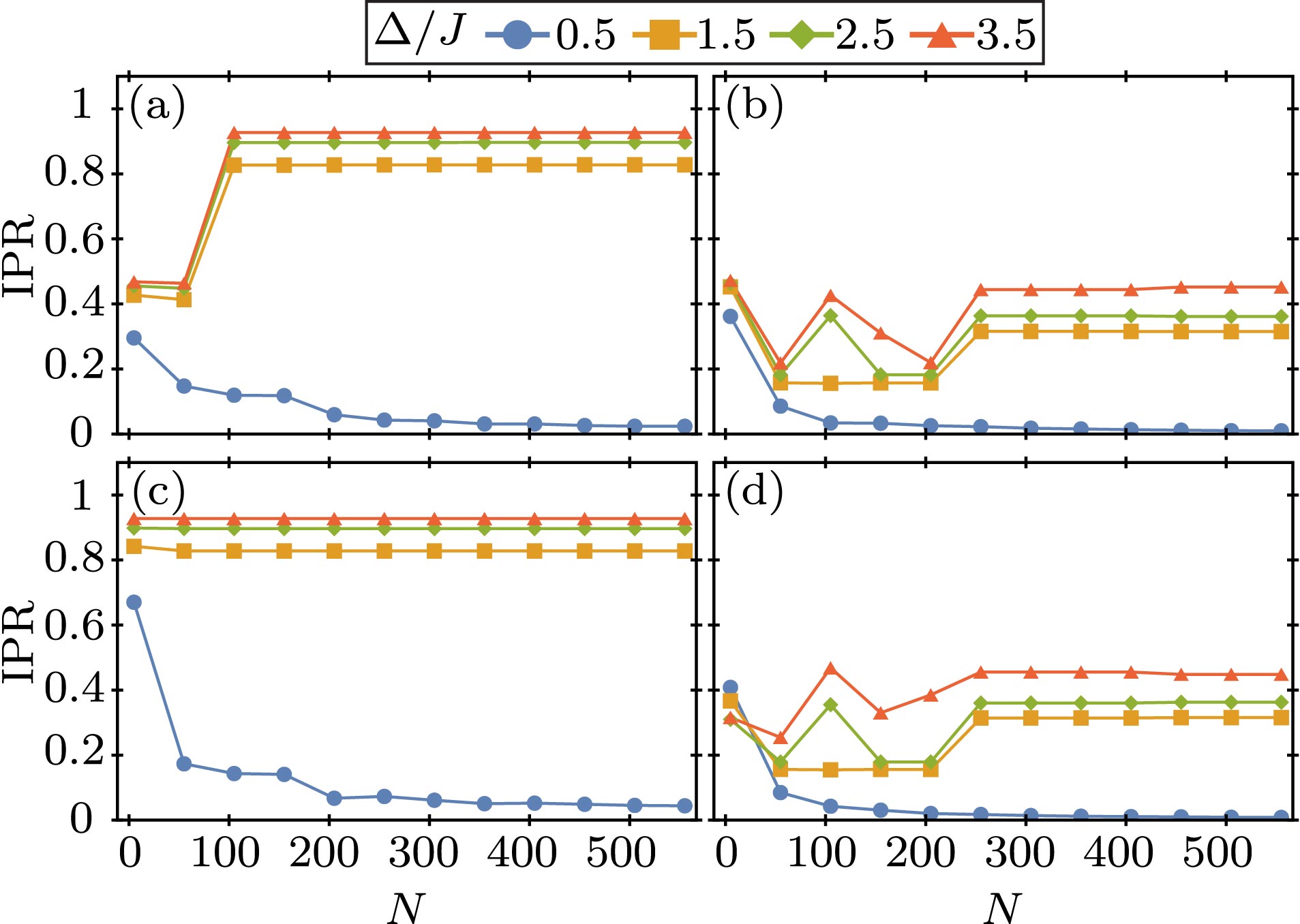}
    \caption{Plots of the IPR versus the system size $N$ for $\beta = 532/738$, $k_T a = \pi/2$, $\Delta/J = 1.5$, and $\varphi = 0$.
    In all panels, we use $h_x/J = \{0.5, 1.5, 2.5, 3.5\}$, where (a) $\nu \approx 0$ (ground state), (b) $\nu \approx 1/2$ (quarter-filled state), (c) $\nu \approx 2$ (highest-energy state), and (d) $\nu \approx 3/2$ (3/4-filled state).}
    \label{fig5}
\end{figure}

\begin{figure}
    \centering
    \includegraphics[width=\linewidth]{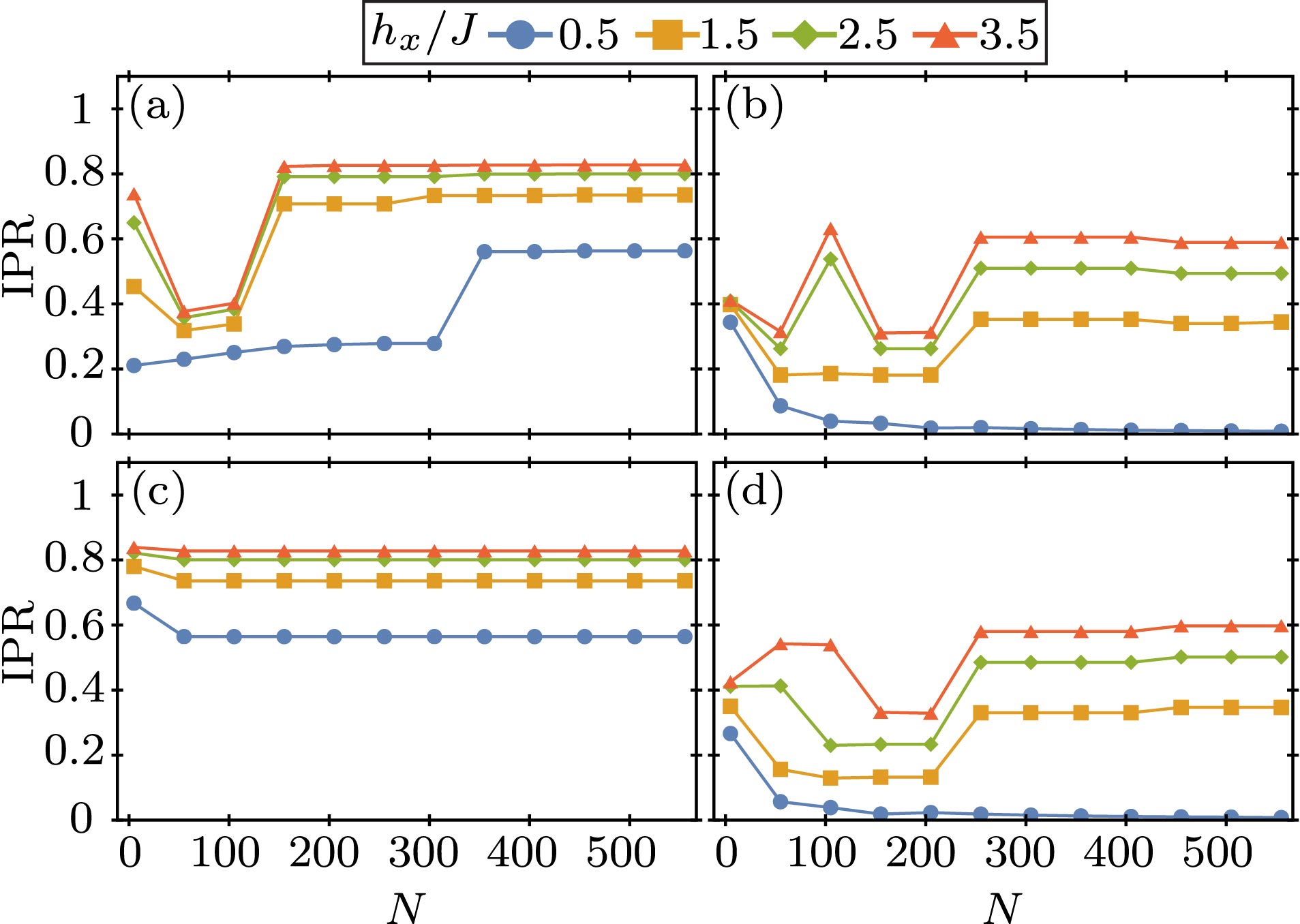}
    \caption{Plots of the IPR versus the system size $N$ for $\beta = 532/738$, $k_T a = 3\pi/8$, $\Delta/J = 1.5$, and $\varphi = 0$.
    In all panels, we use $h_x/J = \{0.5, 1.5, 2.5, 3.5\}$, where (a) $\nu \approx 0$ (ground state), (b) $\nu \approx 1/2$ (quarter-filled state), (c) $\nu \approx 2$ (highest-energy state), and (d) $\nu \approx 3/2$ (3/4-filled state).}
    \label{fig6}
\end{figure}

First, we conduct a scaling analysis of the IPR versus $N$ for a large number of cuts in parameter space, where we fix the filling factor $\nu$ and vary $k_T a$, $h_x/J$, and $\Delta/J$.
The main conclusion of this systematic study is that for $N = 501$, the value used in our exact diagonalization procedure, the IPR has already converged to its thermodynamic limit of infinitely large system size $N \to \infty$ with a relative precision of less than 0.1\%.
In Fig.~\ref{fig4}-\ref{fig6}, we illustrate three examples of the convergence of the IPR as a function of $N$ by the time the system size reaches $N = 501$.
We have tested several examples for larger system sizes $N = 1001$, without any improvements in convergence, showing that the system size chosen for our exact diagonalization procedure is sufficient.

 In Fig.~\ref{fig4} we show plots of the IPR versus system size N for different filling
factors with $\beta = 532/738$, $k_T a = \pi/4$, $h_x/J = 3$, and $\phi = 0$.
In all panels, we use the parameters $\Delta/J = 0.5$ (blue circles), $\Delta/J = 1.5$ (orange squares), $\Delta/J = 2.5$ (green diamonds), and $\Delta/J = 3.5$ (red triangles).
The fillings factors analyzed are $\nu = 0$ (ground state) in Fig.~\ref{fig4}(a), $\nu = 1/2$ (quarter-filled
state) in Fig.~\ref{fig4}(b), $\nu = 2$ (highest-energy state) in Fig.~\ref{fig4}(c),
and $\nu = 3/2$ (3/4-filled state) in Fig.~\ref{fig4}(d).
For localized states, the IPR converges very quickly to asymptotic values at larger values of $N$, while for extended states, the $\rm{IPR} \to 1 /N$ approaches zero asymptotically.
For $N = 501$, the asymptotic limit of $N \to \infty$ is essentially reached.
The converged values of the IPR in Fig.~\ref{fig4}(a) and Fig.~\ref{fig4}(c) are essentially identical, and the same applies to Fig.~\ref{fig4}(b) and Fig.~\ref{fig4}(d), because of the nearly particle-hole symmetric Hamiltonian, which leads to
$\rm{IPR(\nu) \approx \rm{IPR}(2-\nu)}$.

In Fig.~\ref{fig5} we show plots of the IPR versus $N$ for various filling factors $\nu$ for $\beta = 532/738$, $k_T a = \pi/2$, $h_x = 1.5$,
and $\varphi = 0$.
In all panels, we use the parameters $\Delta/J = 0.5$ (blue circles), $\Delta/J = 1.5$ (orange squares), $\Delta/J = 2.5$ (green diamonds), and $\Delta/J = 3.5$ (red triangles).
The filling factors analyzed are $\nu \approx 0$ (ground state) in Fig.~\ref{fig5}(a), $\nu = 1/2$ (quarter-filled state) in Fig.~\ref{fig5}(b), $\nu = 2$ (highest-energy state)
in Fig.~\ref{fig5}(c), and $\nu = 3/2$ (3/4-filled state) in Fig.~\ref{fig5}(d).
For localized states, the IPR converges very quickly to asymptotic values at larger values of $N$, while for extended states, the $\rm{IPR} \to 1/N$ approaches zero asymptotically.
For $N = 501$, the asymptotic limit of $N \to \infty$ is essentially reached.
The converged values of the IPR in Figs.~\ref{fig5}(a) and \ref{fig5}(c) are 
essentially identical, and the same applies to Figs.~\ref{fig5}(b) and \ref{fig5}(d), because of the nearly particle-hole symmetric Hamiltonian, which leads to
$\rm{IPR}(\nu) \ approx \rm{IPR}(2-\nu)$.

In Fig.~\ref{fig6} we show plots
of IPR versus $N$ for $\beta = 532/738$, $k_T a = 3\pi/8$, $\Delta = 1.5$,
and $\varphi = 0$.
In all panels, we use the parameters $h_x/J = 0.5$ (blue circles), $h_x/J = 1.0$ (orange squares), $h_x/J = 1.5$ (green diamonds), and $h_x/J = 2.0$ (red triangles).
The filling factors analyzed are $\nu \approx 0$ (ground state) in Fig.~\ref{fig6}(a), $\nu \approx 1/2$ (quarter-filled state) in Fig.~\ref{fig6}(b), $\nu = 2$ (highest-energy state) in Fig.~\ref{fig6}(c), and $\nu \approx 3/2$ (3/4-filled state) in Fig.~\ref{fig6}(d).
For localized states, the IPR converges very quickly to asymptotic values at larger values of $N$, while for extended states, the
$IPR \to 1/N$ approaches zero asymptotically.
For $N = 501$, the asymptotic limit of $N \to \infty$ is essentially reached.
The converged values of the IPR in Fig.~\ref{fig6}(a) and Fig.~\ref{fig6}(c) are identical, and the same applies to Fig.~\ref{fig6}(b) and Fig.~\ref{fig6}(d),
because of the nearly particle-hole symmetric Hamiltonian, which leads to $\rm{IPR}(nu) \approx \rm{IPR}(2-\nu)$.
The analysis above shows that the IPR for localized states is well converged and that the IPR tends to zero as $1/N$
for extended states.
Thus, for all the ranges of parameters investigated, the number of sites $N = 501$ is essentially in
the thermodynamic limit $(N \to \inf)$.
To reinforce this idea further, we discuss the scaling of the IPR with respect to the system size $L = (N-1)a$.

\subsection{IPR scaling with system size}
An important point in the finite-size scaling analysis is the determination of the IPR’s scaling behavior as a function of the system size $L = (N-1)a$.

\begin{figure}
    \centering
    \includegraphics[width=\linewidth]{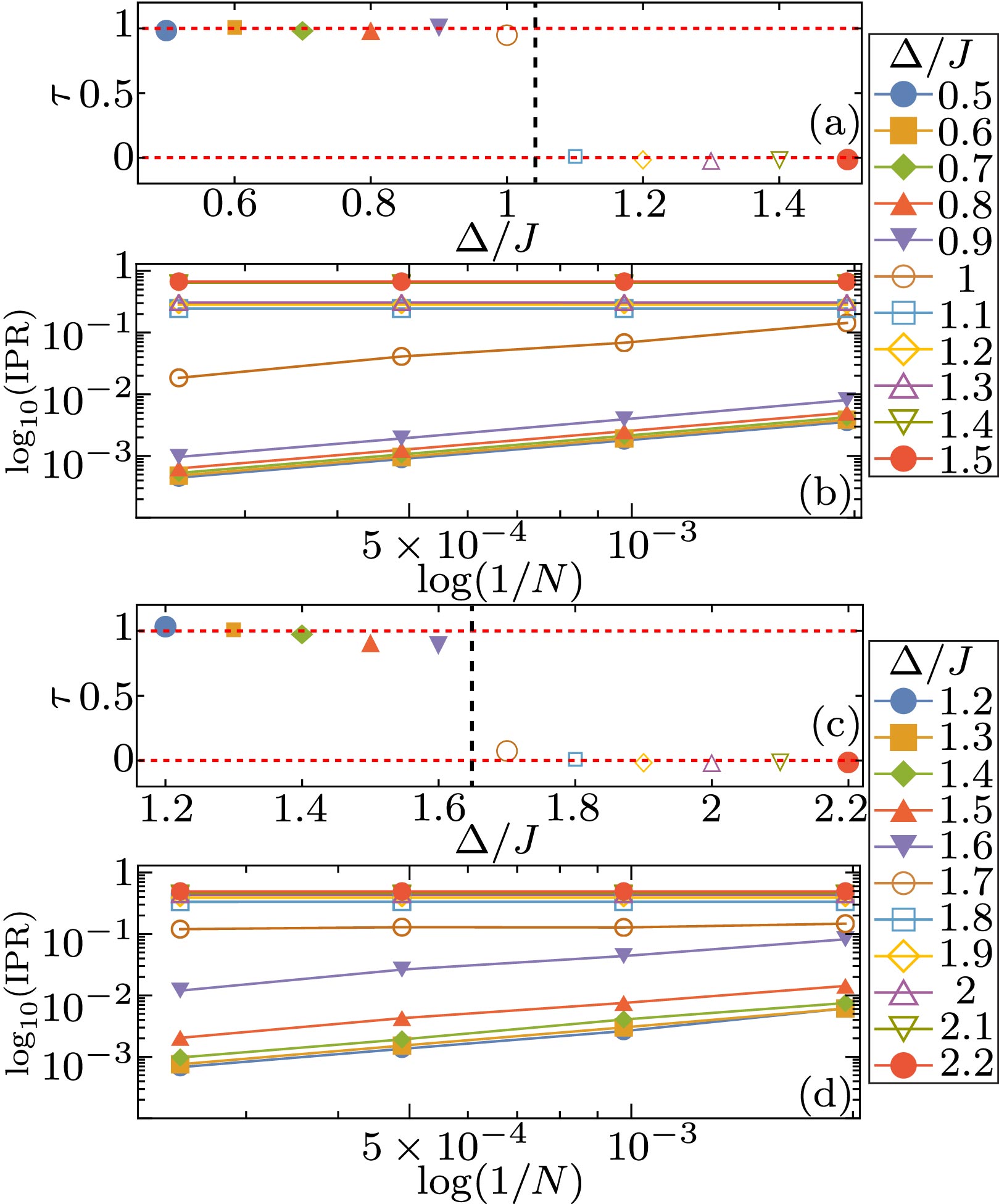}
    \caption{Plots of $\tau$ as a function of $\Delta/J$ and $\log_{10}({\rm IPR})$ versus $\log_{10}(1/N)$ for $\beta = 532/738$, $k_Ta = \pi/4$, $\Delta/J = 3$ and $\varphi = 0$, corresponding to (a) and (b) the ground state $(\nu \approx 0)$ and (c) and (d) the quarted-filled state $(\nu\approx 1/2)$.
    We choose $N = \{512, 1024, 2048, 4096\}$ and the values of $\Delta/J$ are shown in the legends.
    The vertical black dashed lines are the transition points from extended $\tau = 1$ (numerically $\tau \simeq 1$) to localized $\tau = 0$ (numerically $\tau \simeq 0$).}
    \label{fig7}
\end{figure}

\begin{figure}
    \centering
    \includegraphics[width=\linewidth]{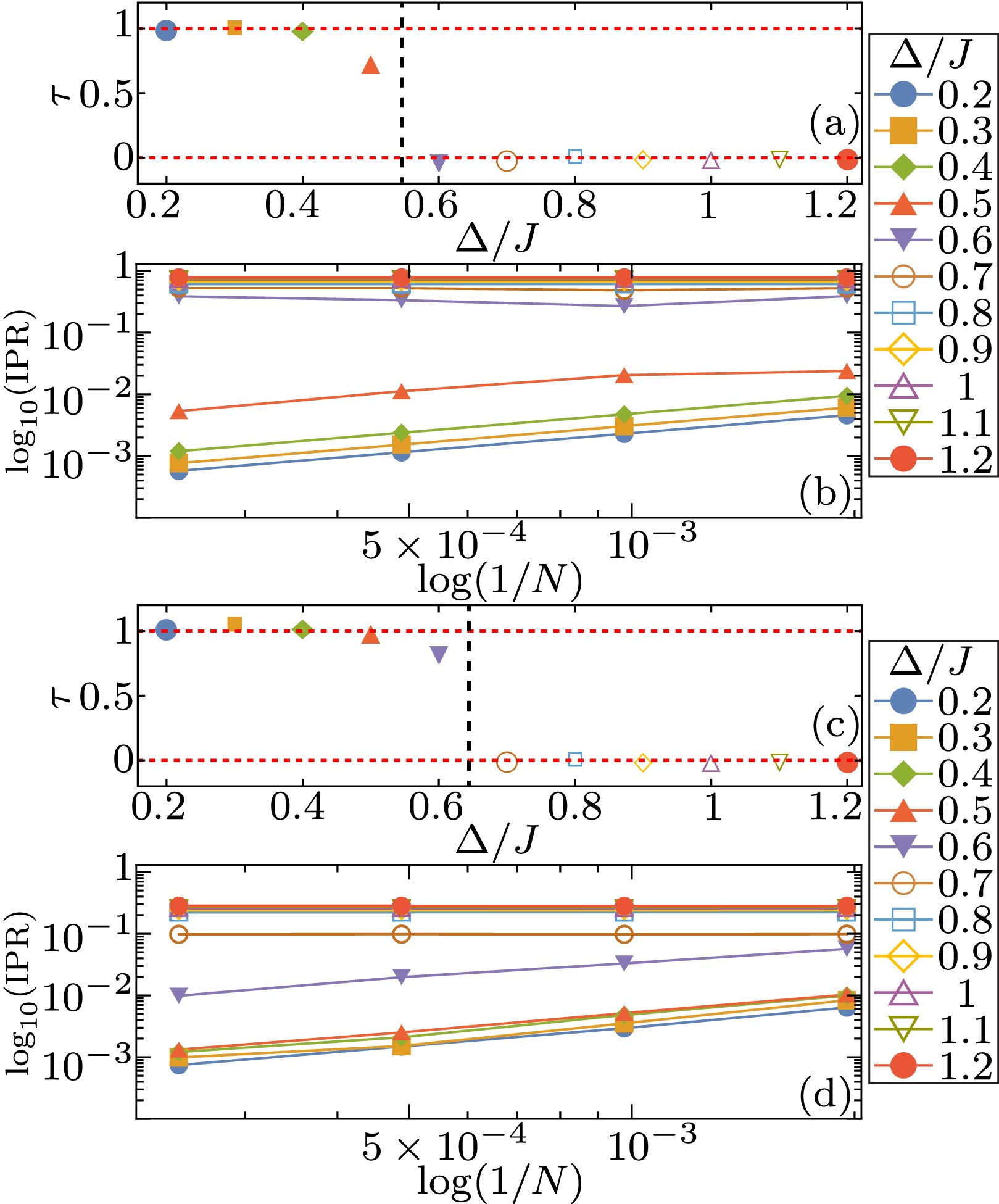}
    \caption{Plots of $\tau$ as a function of $\Delta/J$ and $\log_{10}({\rm IPR})$ versus $\log_{10}(1/N)$ for $\beta = 532/738$, $k_Ta = \pi/2$, $\Delta/J = 1.5$ and $\varphi = 0$, corresponding to (a) and (b) the ground state $(\nu \approx 0)$ and (c) and (d) the quarted-filled state $(\nu\approx 1/2)$.
    We choose $N = \{512, 1024, 2048, 4096\}$ and the values of $\Delta/J$ are shown in the legends.
    The vertical black dashed lines are the transition points from extended $\tau = 1$ (numerically $\tau \simeq 1$) to localized $\tau = 0$ (numerically $\tau \simeq 0$).}
    \label{fig8}
\end{figure}

\begin{figure}
    \centering
    \includegraphics[width=\linewidth]{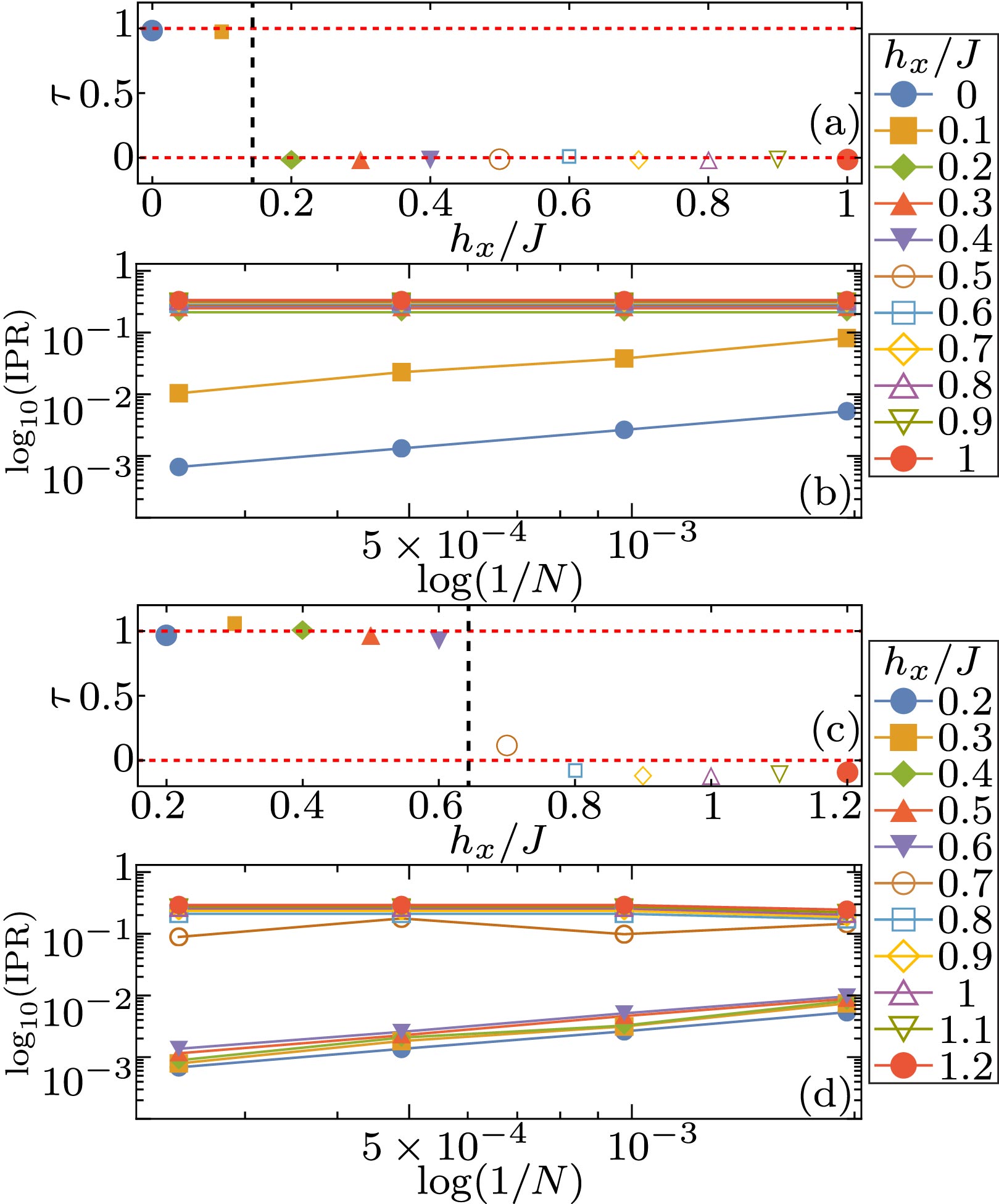}
    \caption{Plots of $\tau$ as a function of $\Delta/J$ and $\log_{10}({\rm IPR})$ versus $\log_{10}(1/N)$ for $\beta = 532/738$, $k_Ta = 3\pi/8$, $\Delta/J = 1.5$ and $\varphi = 0$, corresponding to (a) and (b) the ground state $(\nu \approx 0)$ and (c) and (d) the quarted-filled state $(\nu\approx 1/2)$.
    We choose $N = \{512, 1024, 2048, 4096\}$ and the values of $h_x/J$ are shown in the legends.
    The vertical black dashed lines are the transition points from extended $\tau = 1$ (numerically $\tau \simeq 1$) to localized $\tau = 0$ (numerically $\tau \simeq 0$).}
    \label{fig9}
\end{figure}

As discussed above, in the large-$N$ regime, the IPR behaves as $1/N$ and approaches zero at the thermodynamic limit $(N\to\infty)$ for extended states, while the IPR approaches a nonzero constant when $N \to \infty$ for localized states.
This behavior is best described by the scaling relation \cite{sanchez-palencia-2019}
\begin{equation}
    \rm{IPR} \sim L^{-\tau}.
    \label{eq3.14}
\end{equation}
Where the system is localized when $\tau = 0$ and extended when $\tau = 1$.
This means that by monitoring the exponent
\begin{equation}
    \tau = d \log_{10}(\rm{IPR}) / d \log_{10} (1/N)
    \label{eq3.15}
\end{equation}
for a large number of sites $N$ (large system size $L$), we can determine if the system is localized or extended in the chosen parameter space $\{ \Delta/J, h_x/J, k_T a, \nu \}$.
We chose a few examples in our parameter space to illustrate the behavior of $\tau$ and to reveal the transition point between localized and extended states.
These examples are shown in Figs.~\ref{fig7}-\ref{fig9}.

In Fig.~\ref{fig7} we plot $\tau$ as a function of $\Delta/J$ and $\log_{10}(\rm{IPR})$ versus $log_{10}(1/N)$ for parameters $\beta = 532/738$, $k_T a = \pi/4$, $h_x/J = 3$, and $\varphi = 0$.
These parameters are the same as in Fig.~\ref{fig4}, where we show the convergence of the IPR versus $N$.
Figs.~\ref{fig7}(a) and {fig7}(b) describe the ground state $(\nu\approx 0)$, while Figs.~\ref{fig7}(c) and \ref{fig7}(d) correspond to the quarter-filled state $\nu \approx 1/2$.
We choose $N = {512, 1024, 2048, 4096}$ and the values of $\Delta / J$ are shown in the legends.
The vertical black dashed lines are the transition points from extended $\tau = 1$ (numerically $\tau \simeq 1$) to localized $\tau = 0$ (numerically $\tau \simeq 0$).

In Fig.~\ref{fig8} we show $\tau$ as a function of $\Delta / J$ and $\log_{10}(\rm{IPR})$ versus $log_10(1/N)$ for the parameters $\beta = 532/738$, $k_T a = \pi/2$, $h_x/J = 1.5$, and $\varphi = 0$.
These parameters are the same as in Fig.~\ref{fig5}, where we show the convergence of the IPR versus $N$.
Fig.~\ref{fig8}(a) and Fig.~\ref{fig8}(b) correspond to the ground state $(\nu \approx 0)$, while Fig.~\ref{fig8}(c) and Fig.~\ref{fig8}(d) describe the quarter-filled state ($\nu \approx 1/2$).
We choose $N = {512, 1024, 2048, 4096}$ and the values of $\Delta /J$ are shown in the legends.
The vertical black dashed lines are the transition points from extended $\tau = 1$ (numerically $\tau \simeq 1$) to localized $\tau = 0$ (numerically $\tau \simeq 0$).

In Fig.~\ref{fig9} we plot $\tau$ versus $\Delta / J$ and $\log_{10}(\rm{IPR})$ as a function of $\log_{10}(1/N)$ for the parameters $\beta = 532/738$, $k_T a = 3\pi/8$, $\Delta / J = 1.5$, and $\varphi = 0$.
These parameters are the same as in Fig.~\ref{fig6}, where we show the convergence of the IPR versus $N$.
Figs.~\ref{fig9}(a) and \ref{fig9}(b) correspond to the ground state $(\nu \approx 0)$, while Figs.~\ref{fig9}(c) and \ref{fig9}(d) correspond to the quarter-filled state $(\nu \approx 1/2)$.
We choose $N = \{512, 1024, 2048, 4096\}$ and the values of $h_x/J$ are shown in the legends.
The vertical black dashed lines are the transition points from extended $\tau = 1$ (numerically $\tau \simeq 1$) to localized $\tau = 0$ (numerically $\tau \simeq 0$).
The main messages of Figs.~\ref{fig7}-\ref{fig9} is that the phase boundaries between localized ($\tau = 0$) and extended ($\tau = 1$) states can be systematically determined via this scaling analysis.
To reinforce this idea further, we discuss next the scaling of the IPR with respect to the localization length.

\subsection{IPR scaling with localization length}
We perform an analysis connecting the IPR, the localization length $\xi$, and the system size $L$.
We follow a similar procedure used for the case of localization in Bose-Einstein condensates \cite{sa-de-melo-2023}.
The first step is to establish the relationship between the width $\Delta x$ of an eigenstate spinor $\psi_{is}$ and the
localization length $\xi$.
For this purpose, we define
\begin{equation}
    \braket{\tilde{x}^2} = \sum_{is} i^2 \vert \psi_{is} \vert^2 = \sum_i i^2 \chi_i,
    \label{eq3.16}
\end{equation}
where $\tilde x = x/a$ is the dimensionless position, $\chi_i = \sum_s \vert \psi_{is}\vert^2$ represents the local probability, and the sum over $i$ runs from $-M$ to $+M$, where $M = (N-1)/2$ is a positive integer and $N$ is the number of sites (chosen to be odd for convenience).
We also define 
\begin{equation}
    \braket{\tilde{x}} = \sum_{is} i \vert \psi_{is} \vert^2 = \sum_i i \chi_i
    \label{eq3.17}
\end{equation}
such that the square of the dimensionless width of the eigenstate is
\begin{equation}
    (\Delta \tilde{x})^2 = \braket{\tilde{x}^2} - \braket{\tilde{x}}^2.
    \label{eq3.18}
\end{equation}
For a localized wave function $\psi_{is} = A\exp(-\vert x_i\vert/ \xi)$ around site $j=0$, where $A$ is the normalization constant, the squae width $(\Delta \tilde{x})^2$ can be calculated analytically for any value of the parameters $a/\xi$ and $L/\xi$.
We particulary interested in the regime $a \ll \xi \ll L$, that is $a/\xi \ll 1 \ll L/\xi$, where $L = 2Ma$ is the system size.
Taking the first thermodynamic limit $1\ll L/\xi$, we get the simple analytical form
\begin{equation}
    (\Delta \tilde x)^2 = \frac{1}{2\sinh^2(a/\xi)},
    \label{eq3.19}
\end{equation}
which, for $a/\xi \ll 1$, reduces to
\begin{equation}
    \vert \Delta \tilde x \vert \approx \frac{1}{\sqrt{2}}\left(\frac{\xi}{a}\right),
    \label{eq3.20}
\end{equation}
showing the proportionality of the wave-function width $\Delta \tilde x$ and the localization length $\xi$.

For the localized wave function discussed above, we obtain the expression
\begin{equation}
    {\rm IPR} = \frac{F_M(2\eta)}{\big[ F_M(\eta) \big]^2} ,
    \label{eq3.21}
\end{equation}
where the function is defined as
\begin{equation}
    F_M(\eta) = \coth(\eta) - \frac{e^{-2M\eta} e^{-\eta}}{\sinh(\eta)}.
    \label{eq3.22}
\end{equation}
Here the parameter $\eta = a/\xi$ is the ratio between the lattice spacing $a$ and the localization length $\xi$ and $M$ is a measure of the length of the system $L = 2Ma = (N-1)a$, where $N = 2M +1$ is the number of sites.
Note that $0 \le {\rm IPR} \le 1$.

\begin{figure}
    \centering
    \includegraphics[width=\linewidth]{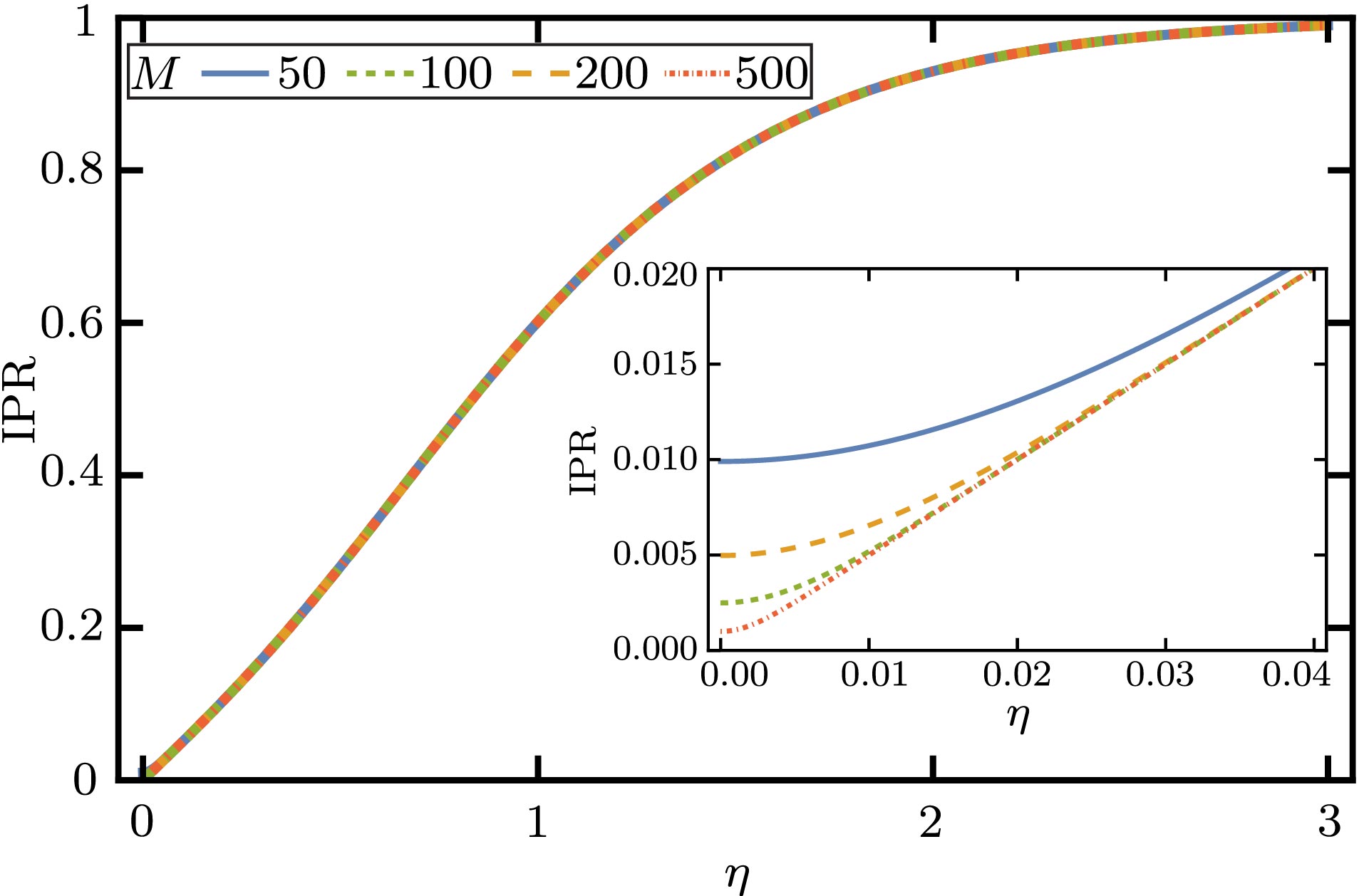}
    \caption{Analytical plots of the IPR versus $\eta$ for different system sizes represented by $M$.
    The inset shows that the IPR tends to $1/N = 1/(2M+1)$ in the limit of $\eta \to 0$ $(\xi/a \to \infty)$ corresponding to a fully extended state.}
    \label{fig10}
\end{figure}

As seen in Fig.~\ref{fig10}, at a fixed number of sites $N=2M+1$, a fully localized state $\eta \to \infty$ (or $\xi/a \to 0$) has ${\rm IPR \to 1}$.
However, at fixed $N$, a fully extended state $\eta \to 0$ (or $\xi/a \to \infty$) has ${\rm IPR} \to 1/N$, as shown in the inset of Fig.~\ref{fig10}.
Furthermore, when $M\to \infty\ (N\to\infty)$, the expression ${\rm IPR} = \coth(2\eta)/\big[ \coth(\eta) \big]^2$ is the thermodynamic limiting curve for the IPR as a function of $\eta$.
An important regime of interest for the IPR given in Eq.~[\ref{eq3.21}] is $a \ll \xi \ll L$, that is, $a/\xi \ll 1 \ll L/\xi$, with $L=2Ma$.
So, taking first the thermodynamic limit $1 \ll L/\xi$, followed by $\eta = a / \xi \ll 1$, leads to ${\rm IPR} \approx \eta / 2 = a / 2\xi$, not shown in the inset of Fig.~\ref{fig10}.

After analyzing the IPR convergence in detail and performing a systematic scaling analysis of its behavior, we found that a system size of $N = 501$ $(M =250)$ is already very close to the thermodynamic limit for the range of parameters investigated.
Thus, next we discuss phase diagrams describing localized and extended states in phase spaces involving the disorder $\Delta / J$, the Rabi field $h_x/J$, and the spin-orbit coupling $k_T a$.

\section{Phase Diagrams}
\label{sec3.5}

\begin{figure}
    \centering
    \includegraphics[width=\linewidth]{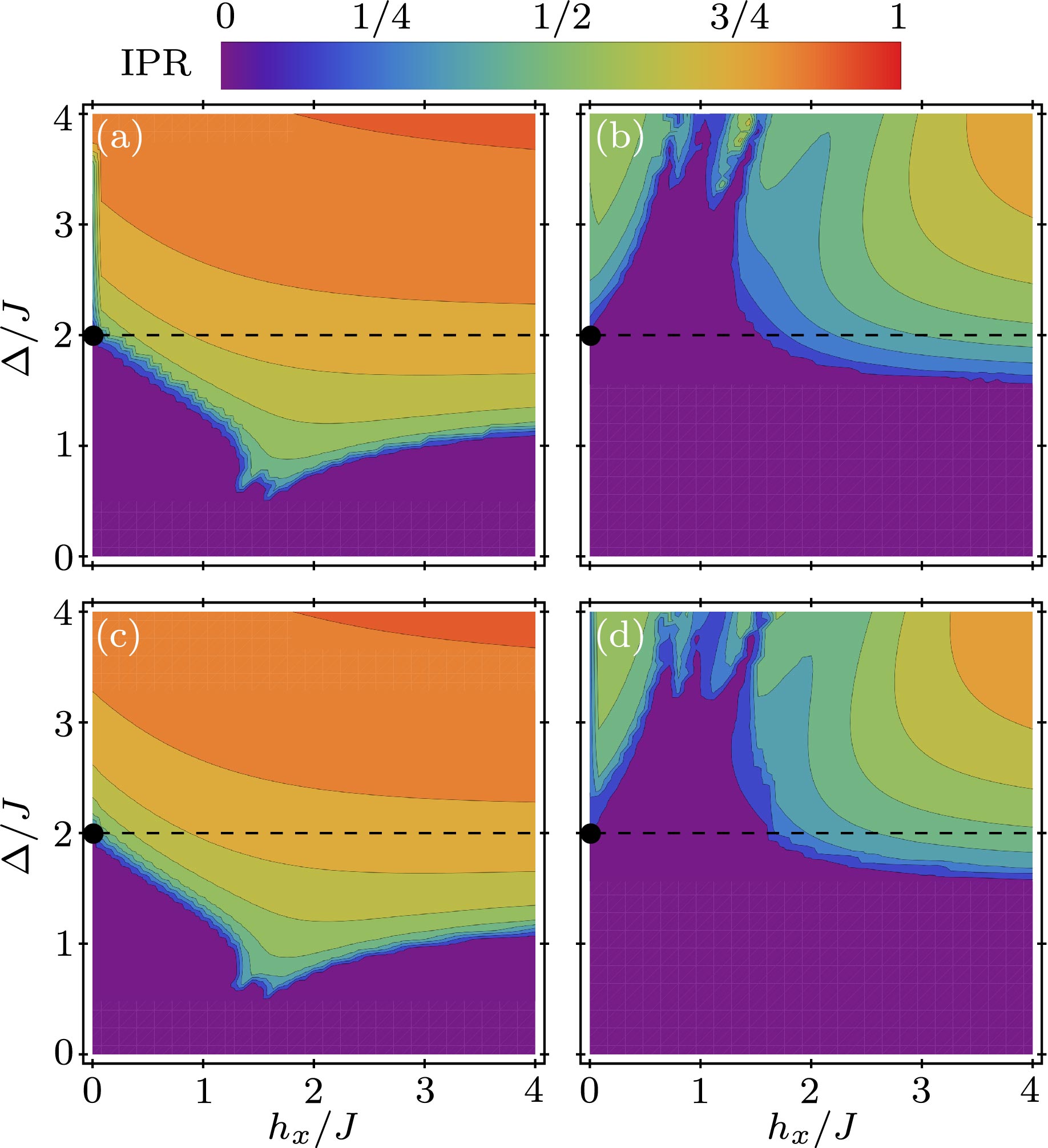}
    \caption{Phase diagrams of $\Delta/J$ versus $h_x/J$ for $\beta = 532/738$, $N = 501$, $k_T a = \pi/4$, $\varphi = 0$, and various filling factors $\nu$:
    (a) $\nu = 1 / 501 \approx 2 \times 10^{-3}$, (b) $\nu = 251 / 501 \approx 1/2$, (c) $\nu = 996/501 \approx 2$, and (d) $\nu = 751 / 501 \approx 3/2.$
    The IPR value changes from fully extended (violet) to fully localized (red).}
    \label{fig11}
\end{figure}

\begin{figure}
    \centering
    \includegraphics[width=\linewidth]{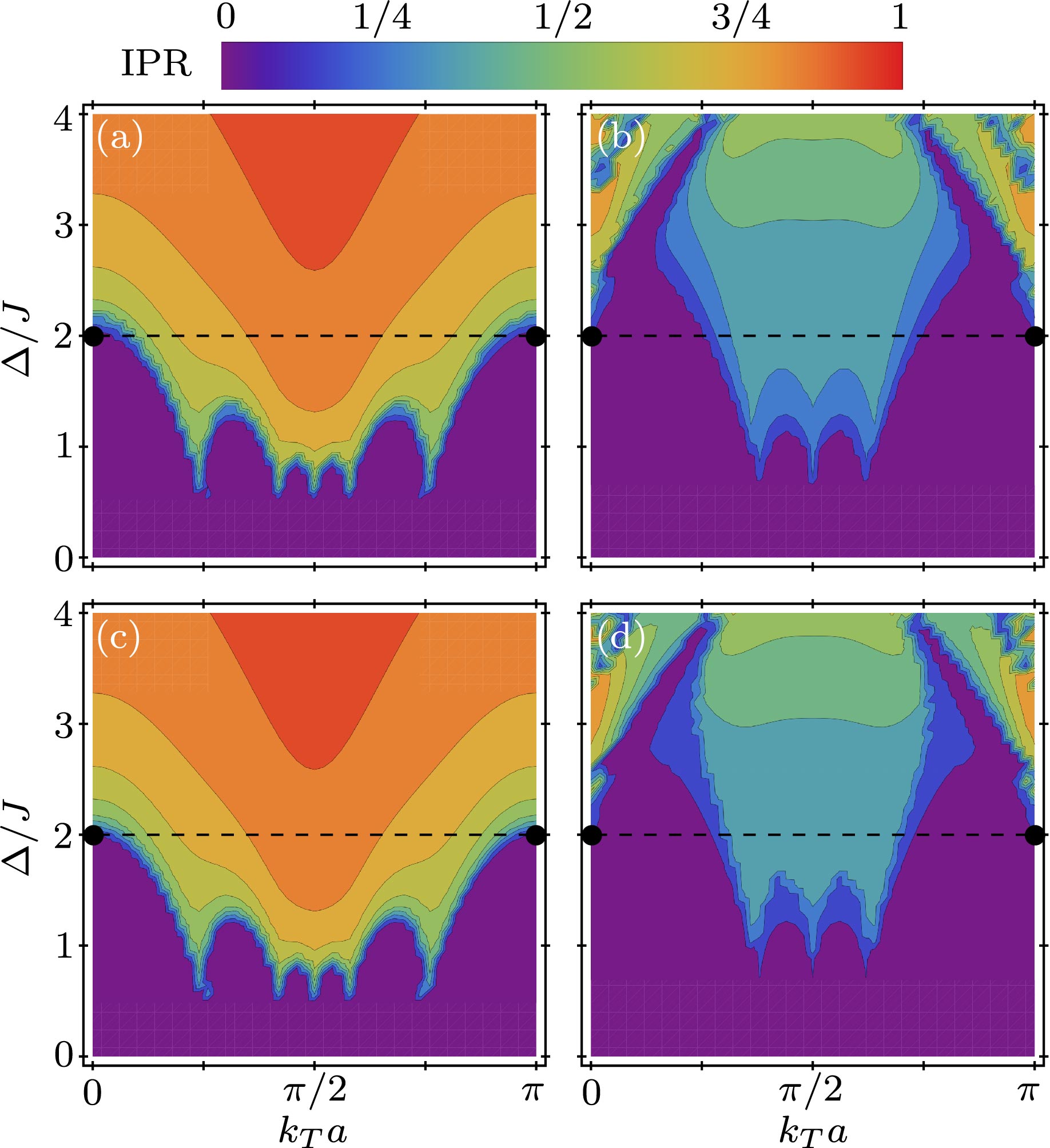}
    \caption{Phase diagrams of $\Delta/J$ versus $k_T a$ for $\beta = 532/738$, $N = 501$, $h_x/J = 1.5$, $\varphi = 0$, and various filling factors $\nu$:
    (a) $\nu = 1 / 501 \approx 2 \times 10^{-3}$, (b) $\nu = 251 / 501 \approx 1/2$, (c) $\nu = 996/501 \approx 2$, and (d) $\nu = 751 / 501 \approx 3/2.$
    The IPR value changes from fully extended (violet) to fully localized (red).}
    \label{fig12}
\end{figure}

\begin{figure}
    \centering
    \includegraphics[width=\linewidth]{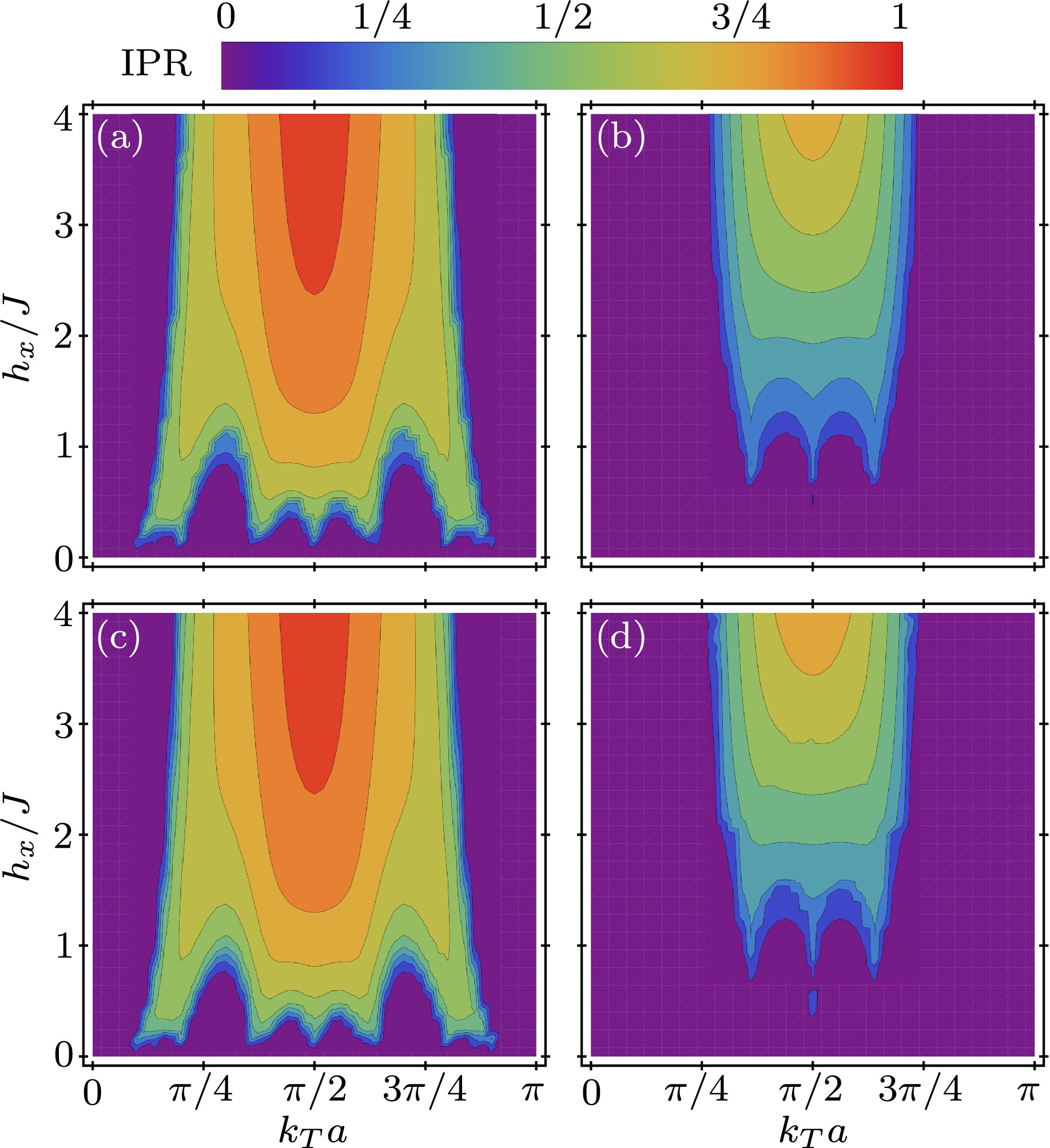}
    \caption{Phase diagrams of $h_x$ versus $k_T a$ for $\beta = 532/738$, $N = 501$, $\Delta = 1.5$, $\varphi = 0$, and various filling factors $\nu$:
    (a) $\nu = 1 / 501 \approx 2 \times 10^{-3}$, (b) $\nu = 251 / 501 \approx 1/2$, (c) $\nu = 996/501 \approx 2$, and (d) $\nu = 751 / 501 \approx 3/2.$
    The IPR value changes from fully extended (violet to fully localized (red).}
    \label{fig13}
\end{figure}

Using the scaling analysis discussed in \ref{sec3.4}, we construct phase diagrams separating regions of localized and extended states.
We verified that the thermodynamic phase boundaries between extended $(\tau = 1)$ and localized $(\tau = 0)$ states are well determined by the numerical boundaries between the violet $\tau \simeq 1$ and nonviolet $(\tau \simeq 0)$ colors in
the planes of $\Delta / J$ versus $h_x/J$ (Fig.~\ref{fig11}), $\Delta/J$ versus $k_T a$ (Fig.~\ref{fig12}), and $h_x/J$ versus $k_T a$ (Fig.~\ref{fig13}), where $N = 501$.
This indicates that for $N = 501$, our system is essentially in the thermodynamic limit for the range of parameters used.
We also checked explicitly that increasing the values of $N$ does not change the numerical phase boundary between the violet (extended, $\tau \simeq 1$) and nonviolet (localized, $\tau \simeq 0$) regions within a relative precision of less than 0.1\%.

In Fig.~\ref{fig11} we show the phase diagram of $\Delta /J$ versus $h_x/J$ for fixed spin-orbit coupling $k_T a$ and changing filling factor $\nu$.
The violet color indicates the region of extended states, where
the IPR tends to zero as $1/N$, with $N$ the number of sites.
The nonviolet (blue to red) colors represent regions of localized states where the IPR converges to a nonzero value.
The parameters are $\beta = 532/738$, $N = 501$, $\varphi = 0$, $k_T a = \pi/4$, and various filling factors $\nu = N_{\rm st}/N$, where ${\rm N}_st$ ($N$) is the number of states (sites).
The range of $\nu$ is from 0 to 2.
This figure displays the phase boundary between extended and localized states.
In all panels, the black dashed line at $\Delta/J = 2$ represents the localization threshold of the Aubry-Andr{\'e} model.
When $h_x/J = 0$, due to spin-gauge symmetry, the critical $(\Delta/J)_c = (\Delta/J)_c^{{\rm AA}}$ in all panels is represented by closed black circle.
There are three main messages in Fig.~\ref{fig11}, which we describe below.

The first message of Fig.~\ref{fig11} is that the phase diagrams of $\Delta / J$ vs $h_x/J$ are nearly particle-hole symmetric, that is, the eigenenergies $E(\nu)$ at filling factors
$\nu$ and $2-\nu$ are nearly symmetric around half filling ($\nu = 1$).
Compare Fig.~\ref{fig11}(a) to Fig.~\ref{fig11}(c) and Fig.~\ref{fig11}(b) to Fig.~\ref{fig11}(d).
The small asymmetries originate from the spatial periods imposed by the quasiperiodic potential and spin-orbit coupling, which, due to the boundary conditions, affect edge states differently for positive and negative energies.
Since the number of edge states is sparse, the energies for the highest occupied states $E(\nu)$, at filling factor $\nu$, and $E(2-\nu)$, at filling factor $2-\nu$, have almost always the same magnitude, leading to an almost symmetric density of states for positive and negative energies.

The second point of Fig.~\ref{fig11} is that for low $\nu \sim 0$ or high $\nu-2$ filling factors [see Fig.~\ref{fig11}(a) and Fig.~\ref{fig11}(c)], the states at $E(\nu)$ can localize (nonviolet regions) as a function of $h_x/J$ at values of $\Delta / J \le 2$, that is, below the localization threshold
of the Aubry-Andr{\'e} model $\Delta/J = 2$, indicated by the black
dashed line.
However, for intermediate filling factors $1/2 \lesssim \nu \lesssim 3/2$ around half filling $(\nu = 1)$, extended states can exist beyond the threshold $\Delta / J = 2$ (black dashed line) depending on $h_x /J$ [see Fig.~\ref{fig11}(b) and Fig.~\ref{fig11}(d)].
This shows that for fixed $k_T a$ and $\beta$, the Rabi field $h_x / J$ can either enhance or hinder localization depending on the filling factor.

The third message of Fig.~\ref{fig11} is that the interplay between $\Delta / J$ and $h_x/J$ for fixed $k_T a$ and $\beta$ may lead to reentrant behaviors.
At low and high filling factors and fixed $\Delta / J \sim 1$ [see Fig.~\ref{fig11}(a) and Fig.~\ref{fig11}(c)], as $h_x/J$ increases, the state at $E(\nu)$ is extended for $0 \leq h_x/J \lesssim 1.5$, then localizes for $1.5 \lesssim h_x/J \lesssim 2$, and then is extended again for $h_x/J \gtrsim 2$.
This reentrant behavior is due to the helical structure of the
spinor state at $E(\nu)$, which reduces the effective hopping
between neighboring sites and thus causes localization in the
region $1.5 \lesssim h_x/J \lesssim 2.0$.
At intermediate filling $1/2 \lesssim \nu \lesssim 3/2$ and fixed $\Delta / J = 2.5$ [see Fig.~\ref{fig11}(b) and Fig.~\ref{fig11}(d)], localized states occur for $0 \lesssim h_x/J \lesssim 0.1$, extended states arise for $0.1 \lesssim h_x/J \lesssim 1.5$, and localized states arise again for $h_x / J \gtrsim 1.5$. This reentrant behavior at $\Delta / J = 2.5$ is also due to the helical structure of the spinor state at $E(\nu)$, which now enhances the effective hopping between neighboring sites and thus produces extended states in the region $0.1 \lesssim h_x / J \lesssim 1.5$.

In Fig.~\ref{fig12} we show the phase diagram of $\Delta/J$ versus $k_T a$ for fixed Rabi field $h_x/J$ and changing filling factor $\nu$.
The violet color indicates the region of extended states, where the IPR tends to zero as $1/N$, with $N$ the number of sites.
The nonviolet (blue to red) colors represent regions of localized states where the IPR converges to a nonzero value.
The parameters used are $\beta = 532/738$, $N = 501$, $\varphi = 0$, $h_x/J = 1.5$, and various filling factors $\nu = N_{\rm st}/N$, where $N_{\rm st}$ ($N$) is the number of states (sites).
The range of $\nu$ is from 0 to 2.
This figure displays the phase boundary between extended and localized states, the $\pi$ periodicity of the phase diagram, and the reflection symmetry around $k_T a = \pi/2$.
In all panels, the black dashed line at $\Delta / J = 2$ represents the localization threshold of the Aubry-Aubry-Andr{\'e}.
Notice that $(\Delta / J)_c = (\Delta / J)_c^{\rm AA} = 2$ when $h_x / J = 0$ due to the spin-gauge symmetry; the black dots indicate this transition at $k_T = 0$ and $\pi$.
The finite Rabi field makes two independent copies of the Aubry-Andr{\'e} model by shifting the energy spectrum of up and down spins but maintaining the critical value for the transition between extended and localized states.
The interplay between local energy (controlled by $\Delta$) and the local fields (controlled by $h_x$) in Eq.~[\ref{eq3.6}] gives rise to the periodic fingering phenomenon with the largest localization value at $k_T a = \pi /2 ({\rm mod}\pi)$ for sufficient $h_x/J$.

The panels in Fig.~\ref{fig12} have nearly particle-hole symmetry; compare Fig.~\ref{fig12}(a) with $\nu \approx 0$ and Fig.~\ref{fig12}(b) with $\nu \approx 2$, as well as Fig.~\ref{fig12}(c) with $\nu \approx 1/2$ and Fig.~\ref{fig12}(d) with $\nu \approx 3/2$.
In Fig.~\ref{fig12}(a) and Fig.~\ref{fig12}(c), for low and high filling factors, there are localized states (nonviolet regions) for $\Delta / J <2$ due to reduced mobility regions.
In Fig.~\ref{fig12}(b) and Fig.~\ref{fig12}(d), for intermediate filling factors, there are localized states (nonviolet regions) for $\Delta / J < 2$, but also extended states (violet regions)
for $\Delta / J > 2$ (violet regions).
The fingering phenomenon in the panels is the result of the interplay between the local energies $\Delta \cos(2\pi \beta i)$ and the local fields $\tilde h_{\perp} = h_x e^{-{\rm i}2k_T a i}$. 
Like the Hamiltonian $\mathcal H_{\varphi}$ Eq.~[\ref{eq3.1}], the IPR is a periodic function of $k_T a$ with period $\pi$, reaching larger values for $k_T a = \pi/2 \ {\rm mod}\pi$ when $h_x / J$ is sufficiently large.
The symmetry line at $k_T a = \pi/2$ arises when the site-dependent complex Rabi field $\tilde h_\perp = \tilde h_x - i \tilde h_y$ is staggered, with $\tilde h_x = h_x(-1)^i$ and
$\tilde h_y = 0$, producing spin inhomogeneity that facilitates localization.
In contrast, along the symmetry line $k_T a = 0\ ({\rm mod}\pi)$
the spin inhomogeneity is absent since $\tilde h_\perp$ is uniform, with $\tilde h_x = h_x$ and $\tilde h_y = 0$, thus facilitating delocalization.

In Fig.~\ref{fig13} we show phase diagrams of $h_x/J$ versus $k_T a$ for fixed $\Delta / J =1.5$ and the same filling factors as in Fig.~\ref{fig12}.
In the absence of $h_x/J$, where the spin-gauge symmetry is
present and the Aubry-Andr{\'e} model is recovered, it is evident
that all states are extended since $\Delta / J < 2$. Furthermore, when $k_T a = 0 ({\rm mod}\pi)$ but $h_x/J \ne 0$, there are two copies of the Aubry-Andr{\'e} model, and all states are again extended since $\Delta / J \le 2$. 
The same fingering phenomenon, like in Fig.~\ref{fig12}, emerges as $k_T a$ changes, and the same periodic behavior as a
function of $k_T a$ arises.
The small little islands in Figs.~\ref{fig13}(b) and \ref{fig13}(d) reflect localized edge states.

\begin{figure}
    \centering
    \includegraphics[width=\linewidth]{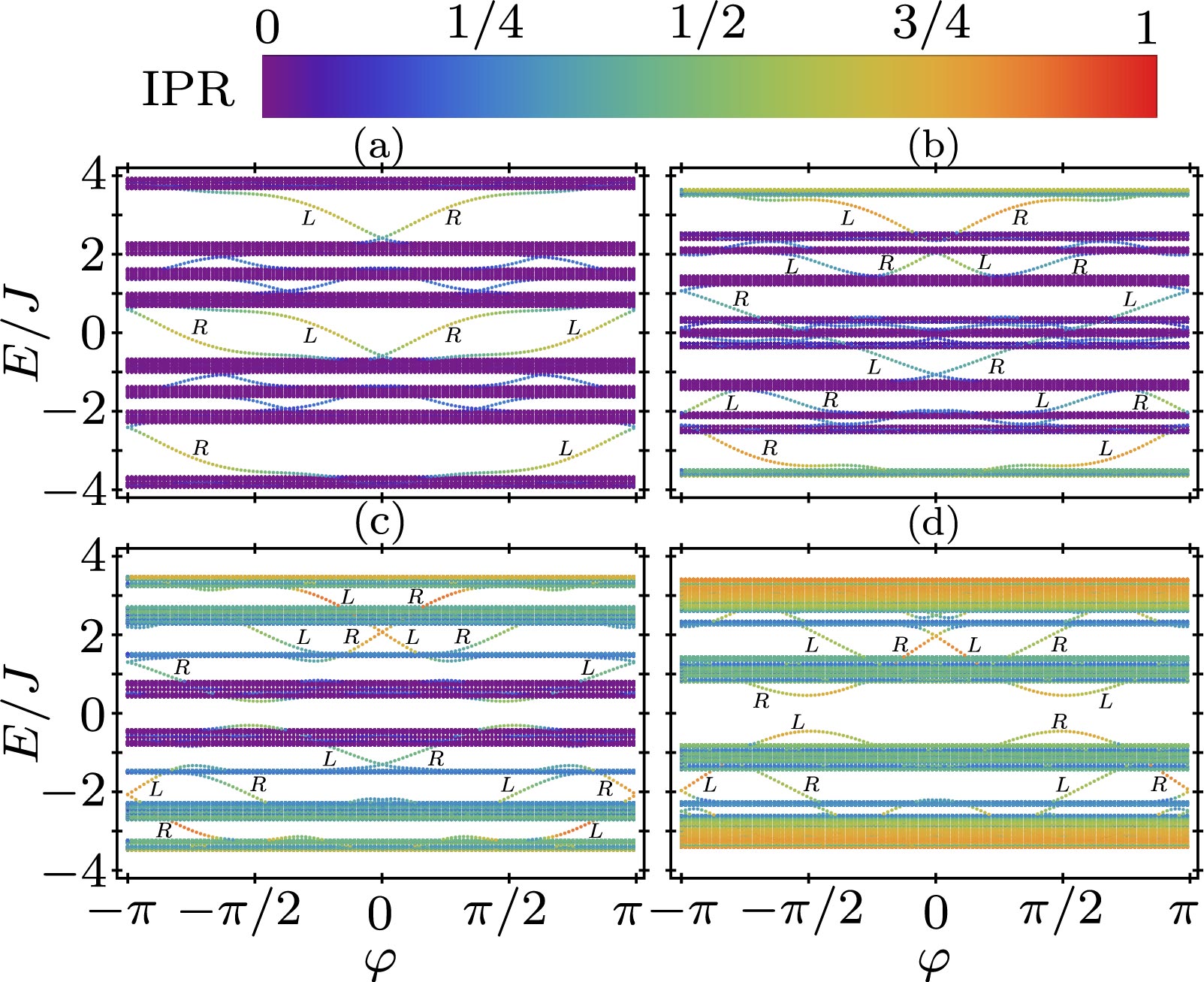}
    \caption{Plots of $E/J$ versus $\varphi$ with $\beta = 532/738$, $\Delta/J = 1.5$, $h_x = 1.5$, $N = 501$, and fixed values of the SOC parameter: (a) $k_T a = 0$, (b) $k_T a = \pi/4$, (c) $k_T a = 3\pi/8$, and (d) $k_T a = \pi/2$.
    The lines in the bulk gaps represent edge states labeled $R$ (right) and $L$ (left); chiral edge states cross between bulk bands.}
    \label{fig14}
\end{figure}
\section{Topological Properties}
\label{sec3.6}

In Fig.~\ref{fig14} we show plots of energy $E/J$ versus $\varphi$, indicating the IPR on a color scale for $\beta = 532/738$, $\Delta/J = 1.5$, $h_x/J = 1.5$, and system size of $N = 501$ for different values of SOC: $k_T a = 0$ [Fig.~\ref{fig14}(a)], $k_T a = \pi/4$ [Fig.~\ref{fig14}(b)], $k_T a = 3\pi/8$ [Fig.~\ref{fig14}(c)], and $k_T a = \pi/2$ [Fig.~\ref{fig14}(d)].
These parameters are the same as those of the vertical dashed lines shown in Fig.~\ref{fig2}.
The violet color indicates fully extended states and the red color describes fully localized states.
The criterion for localization is established via the finite-size scaling analysis discussed in \ref{sec3.4}.
We calculate the IPR of each state, defined in Eq.~[\ref{eq3.11}], and label the states as extended or localized to visualize the effects of $h_x/J$ on localization.
In Fig.~\ref{fig14}(a), where $k_T a = 0$, the system remains self-dual for any value of $h_x/J$, that is, a global SU(2) rotation creates two copies of the AA model with different energy references for each spin component.
Thus, all bulk states are extended (violet) for $\Delta / J = 1.5$, since $\Delta / J < (\Delta / J)_c^{\rm AA}$.
In this special case, the energies of all states change linearly with $h_x/J$.
In Fig.~\ref{fig14}(b)–Fig.~\ref{fig14}(d), where $k_T a \ne 0$, self-duality is globally broken, allowing for the emergence of mobility regions.
Nonzero values of $k_T a$ create a spin-dependent chiral modulation via the term
\begin{equation}
    \mathbf h (x_n) = -h_x \cos(2k_T x_n) \boldsymbol{\sigma}_x - h_x \sin(2k_T x_n)\boldsymbol{\sigma}_y
    \label{eq3.22}
\end{equation}
which competes with the spatial modulation $\Delta \cos(2\pi \beta n - \varphi) \mathbf I$ controlled by $\Delta$.
Changing $k_T a$ modifies the period of $\mathbf h(x_n)$ affecting the lower- and higher-energy (filling) bulk states the most.
These examples demonstrate the existence of mobility regions and the emergence of localization of bulk states below the critical self-dual value for the AA model.

In general, bulk states near zero energy (half filling), with $k_T a \ne 0$ and $h_x \ne 0$, are the most robust against localization, remaining extended for values of $\Delta / J$ larger than $(\Delta/J)_c^{\rm AA}$.
In addition to bulk states, there are also edge states that
crossover from one band to another.
Edge states are important for topological Anderson insulators in two dimensions, that is, systems that are Anderson localized in the bulk but conducting at the edge \cite{beenakker-2009, shen-2009}.
Edge states are also important for quantum Hall insulators \cite{thouless-1982,kohmoto-1985}.
In our 1D system the existence of these edge states also signals topological phases of matter, characterized by Chern numbers defined in $(1\,+\,1)$-dimensions, where one dimension is real ($x$ direction) and the other is synthetic, defined by $\varphi$.

\subsection{Connection between topological Aubry-Andr{\'e} and quantum Hall insulators}
The connection between the AA Hamiltonian in one dimension to fermions in 2D square lattices with a perpendicular magnetic field can be understood through the Harper-Hofstadter (HH) model \cite{harper-1955, hofstadter-1976}, which was recently simulated with ultracold bosons $({}^{87} {\rm Rb})$ \cite{spielman-2021,zhou-2021}.
The addition of SOC and Rabi fields to the HH model leads to the Hamiltonian
\begin{equation}
    \hat H = \varepsilon(\hat {\mathbf {k}})\mathbf I - h_x \cos(2k_T x_n)\boldsymbol{\sigma}_x - h_x \sin(2k_T x_n)\boldsymbol{\sigma}_y
    \label{eq3.24}
\end{equation}
in a 2D square lattice.
The kinetic energy operator is 
\begin{equation}
    \varepsilon(\hat{\mathbf k}) = -2t_x \cos(\hat {k}_x a_x) -2t_y \cos(\hat {k}_y a_y),
    \label{eq3.25}
\end{equation}
where $\hat{k}_x = -i \frac{\partial}{\partial x}$ and $\hat{k}_y = -i \frac{\partial}{\partial y}$.
When the magnetic field $\mathbf B$ is applied in the $z$ direction, the vector potential can be chosen as $\mathbf A = (0,Bx,0)$.
Shifting $\hat{k}_y \to \hat{k}_y - qA/\hbar c$, where $q = e > 0$ is the charge, the $\cos(\hat{k}_y a_y)$ term becomes $\cos(\hat k_y a_y - eBxa_y/\hbar c)$.
Defining $x = na_x$, we write $eBxa_y/\hbar c = 2\pi n \Phi/\Phi_0$, where $\Phi = B a_x a_y$ is the magnetic flux through the rectangular plaquette and $\Phi_0 = hc/e$ is the flux quantum (cgs units).
Since $k_y$ is conserved, a Fourier transformation to real space generates a family of 1D Hamiltonians labeled by $k_y$,
\begin{equation}
    \hat{H}_{k_y} = -t_x \sum_{\braket{nm}s} f_{nk_ys}^{\dagger} f_{mk_ys} + \sum_{nss'} \Lambda_{nn}^{ss'} f_{nk_ys}^{\dagger} f_{nk_ys'}
    \label{eq3.26}
\end{equation}
where $f_{nk_y s}^\dagger$ creates a fermion at site $n$ with momentum $k_y$ and spin $s$.
The first term is the hopping in the $x$ direction and the second is the modulation along $x$ controlled by the matrix
\begin{multline}
    \boldsymbol{\Lambda}_{nn} = -2t_y \cos(2\pi n \Phi/\Phi_0- k_y a_y)\mathbf I - h_x \cos(2k_T x_n)\boldsymbol{\sigma}_x \\ - h_x \sin(2k_T x_n)\boldsymbol{\sigma}_y.
    \label{eq3.27}
\end{multline}
Making the identification of $\Phi / \Phi_0 \to \beta$, $k_y a_y \to \varphi$, $t_x \to J$, $-2t_y \to \Delta$, $f_{nk_y s}^{\dagger} \to \tilde{c}_{ns}^{\dagger}$, and $\boldsymbol{\Lambda}_{nn} \to \tilde{\boldsymbol{\Gamma}}_{nn}$ produces the Hamiltonian $\tilde {\mathcal H}_{\varphi}$ defined in Eq.~[\ref{eq3.4}].
Therefore, considering $\varphi$ to be a synthetic dimension (equivalent to $k_y$), we have shown an exact mapping of the 2D HH model with SOC and Rabi fields to the 1D family of AA Hamiltonians shown in Eq.~[\ref{eq3.4}].
Naturally, the quantum Hall insulating phases and topological invariants of the HH model in two dimensions also have corresponding partners for the AA model in $1\ +\ 1$ dimensions, one real and one synthetic dimension.

In addition to duality breaking and the existence of mobility regions, the AA model with SOC and Rabi fields also exhibits edge states and unconventional topological insulating phases.
To elucidate the topological nature of these edge states, we can take a second look at Fig.~\ref{fig14}, which contains plots of energy $E/J$ versus $\varphi$, the synthetic momentum dimension.
Extended states appear in violet and nonextended states in other colors.
The energy bands $E(\varphi)$ for bulk states are quite flat in the range $-\pi < \varphi < \pi$, but edge states labeled by $R$ (at right) and $L$ (at left) disperse with $2\pi$ periodicity.
The $R$ and $L$ states indicated can have a spin projection (generalized helicity) $m_s = \pm$.

In Fig.~\ref{fig14}(a) $(k_T a = 0)$, all bulk states are extended, and nontrivial gapped regions represent conventional topological insulators, where edge states merge into extended ones at the low- and high-energy ends of the gap.
In Fig.~\ref{fig14}(b) and \ref{fig14}(c) $(k_T a = \pi/4, 3\pi/8)$ there are mobility regions in the bulk and a new type of phase emerges, the unconventional (hybrid) topological Aubry-Andr{\'e} insulator, where edge states, within nontrivial gapped regions, merge into an extended bulk state starting from a localized bulk state or vice versa.
In Fig.~\ref{fig14}(d) $(k_T a = \pi/2)$ all bulk states are localized, and nontrivial gapped regions represent conventional topological Aubry-Andr{\'e} insulators, where edge states merge into localized bulk states at the low- and high-energy ends of the gap.

\subsection{Edge states and Chern number}
Edge states $R$ and $L$ edge can cross from one band to the other and exhibit helicity in the $(1\,+\,1)$-dimensional space described above.
Defining $\tilde y$ as the dual variable to $k_{\tilde y} = \varphi /a_{\tilde y}$, the geometry of the $x\tilde y$ space is that of a cylinder with finite size along the real space direction $x$ and periodic boundary conditions along the synthetic dual dimension $\tilde y$.
This establishes the bulk-edge correspondence between the total chirality of the edge states in a given bulk gap and the corresponding Chern number.
In a toroidal compactification of our $(1\,+\,1)$-dimensional
space, the Chern number is $C = \sum_{m_s}^{E < \mu} C_{m_s}$, where $C_{m_s}$ is the Chern index
\begin{equation}
    C_{m_s} = \int_{\partial \Omega} dk_x dk_{\tilde y} \mathcal B_{x\tilde y}^{m_s} (k_x, k_{\tilde y}),
    \label{eq3.28}
\end{equation}
with $\mathcal B_{x\tilde y}$ the Berry curvature \cite{kohmoto-1985} for a given spin projection (generalized helicity) $m_s$.
In our $(1\,+\,1)$-dimensional space, $k_x$ is the dual to $x$ and $k_{\tilde y}$ is the dual to $\tilde y$.
As a result, the Wannier-Claro gap labeling theorem \cite{wannier-1978, wannier-1979} for quantum Hall insulators holds, at all gaps shown in Fig.~\ref{fig14}, when spin-orbit coupling and Rabi fields are present \cite{sa-de-melo-2019}
\begin{equation}
    \nu = S + C \frac{\Phi}{\Phi_0},
    \label{eq3.29}
\end{equation}
where $\beta = \Phi/\Phi_0 = 532/738$, $C$ is the Chern number, and $S$ is the topological invariant associated with charge screening of the lattice potential, satisfying the relation
\begin{equation}
    S = n_{\rm ind} a_x a_{\tilde y} = n_{\rm ind}^x a_x.
    \label{eq3.30}
\end{equation}
Here $n_{\rm ind}$ is the induced particle density $a_x \ (a_{\tilde y})$ is the lattice spacing along $x \ (\tilde y)$, or more conveniently $n_{\rm ind}^x$ is the induced particle density per unit length in the $x$ direction.
In \ref{ch3table1} we show $S$ and $C$ within visible bulk gaps labeled $g_1,\dots,g_7$ (from lowest to highest energies), seen in Fig.~\ref{fig14}(c).
The definition and values of $\chi$ are discussed later in this section.

\begin{table}[tb]
    \centering
    \begin{tabular}{@{}cccccccc@{}}
        \toprule \toprule
        \diagbox{Parameter}{$\rm g$} 
        & ${\rm g}_1$ & ${\rm g}_2$ & ${\rm g}_3$ & ${\rm g}_4$ & ${\rm g}_5$ & ${\rm g}_6$ & ${\rm g}_7$ \\ \midrule
        $\nu$  & 0.28 & 0.56 & 0.72 & 1.00 & 1.28 & 1.44 & 1.72 \\
        $S$    & 1.00 & 2.00 & 0.00 & 1.00 & 2.00 & 0.00 & 1.00 \\
        $C$    & -1   & -2   & 1    & 0    & -1   & 2    & 1    \\
        $\chi$ & -1   & 0    & 1    & 0    & 1    & 0    & -1   \\ \bottomrule \bottomrule
    \end{tabular}
    \caption{Values of the filling factor $\nu$, Chern number $\mathcal C$, and quantization of $S$ representing the charge screening for the visible energy gaps in Fig.~\ref{fig14}(c).
    The lowest-energy gap is ${\rm g}_1$ and the highest is ${\rm g}_7$.
    The topological invariant $\mathcal C (\chi)$ represents the charge-charge Chern number $\mathcal C_{\rm ch}^{\rm ch}$  (spin-charge Chern number $\mathcal C_{\rm sp}^{\rm ch}$) obtained from the Chern matrix.}
    \label{ch3table1}
\end{table}

Next, we introduce the concept of the Chern matrix to identify the topological invariants of the insulating phases.
For this purpose, we apply phase twists to the boundary conditions of the many-particle wave function as $\Psi (\mathbf r_{1s}, \dots ,\Psi (\mathbf r_{js} + \mathbf L_\ell, \dots, \mathbf r_{N_ps}) = e^{i\phi_{\ell s}} \Psi(\mathbf r_{1s}, \dots, \mathbf r_{js}, \dots, \mathbf r_{N_p s})$, where $\mathbf r_{js}$ is the position of the $j$th particle of the spin $s$, $\mathbf L_\ell = M_\ell a_\ell \hat{\boldsymbol{\ell}}$ is the length vector in the $\ell = \{x,y\}$ direction, $\phi_{\ell}^{s}$ is the phase twist \cite{wu-1985} along $\ell$ for spin $s$ and $N_p$ is the total number of paticles,

The twisted wave function
\begin{multline*}
    \tilde \Psi (\mathbf r_1s, \dots, \mathbf r_{js}, \dots, \mathbf r_{N_ps}) \\ = \exp \left[ -i \sum_{j,s} \left( \phi_{xs} \frac{x_{js}}{M_x a_x} + \phi_{ys} \frac{y_{js}}{M_y a_y} \right) \right]  \\ \times \Psi (\mathbf r_1s, \dots, \mathbf r_{js}, \dots, \mathbf r_{N_ps}),
\end{multline*}
where $0 \le \phi_{\ell s} < 2\pi$.
Note that $\tilde \Psi (\mathbf r_{1s}, \dots, \mathbf r_{js}, \dots, \mathbf r_{N_p s})$ is periodic in $\phi_{\ell s}$ and that a Chern matrix can be defined as \cite{sheng-2003}
\begin{equation}
    C_{ss'} = \frac{i}{4\pi}\int\int d\phi_{xs} d\phi_{ys'} \mathcal F_{xy}(\phi_{xs},\phi_{ys'}),
    \label{eq3.31}
\end{equation}
where the curvature function does not have a real component, being purely imaginary
\begin{equation}
    \mathcal F_{xy}(\phi_{xs},\phi_{ys'}) = 
    \bigg{\langle} \frac{\partial \tilde{\Psi}}{\partial \phi_{xs}} \bigg{\vert} \frac{\partial \tilde{\Psi}}{\partial \phi_{ys'}} \bigg{\rangle} - 
    \bigg{\langle} \frac{\partial \tilde{\Psi}}{\partial \phi_{ys'}} \bigg{\vert} \frac{\partial \tilde{\Psi}}{\partial \phi_{xs}} \bigg{\rangle}.
    \label{eq3.32}
\end{equation}

The curvature function is integrated over the torus $\mathcal T_{ss'}^2$, i.e., integrated over the phase twists $0 \le \phi_{xs} < 2\pi$ and $0 \le \phi_{ys'} < 2\pi$.
The Chern matrix is a square $2 \times 2$ matrix, since there are two spin (generalized helicity) states.
The expression given in Eq.~[\ref{eq3.31}] is an integer \cite{sa-de-melo-2021} just like in standard SU(2) systems without spin-orbit coupling and Rabi fields \cite{haldane-2006}.

From the Chern matrix in Eq.~[\ref{eq3.31}], three topological invariants can be extracted.
The charge-charge (charge-Hall) Chern number $\mathcal C_{\rm ch}^{\rm ch} = \sum_{cc'} \mathcal C_{cc'}$ is the first topological invariant, while the second is the spin-charge (spin-Hall) Chern number $\mathcal C_{\rm sp}^{\rm ch} = \sum_{ss'} m_s \mathcal C_{ss'}$ or charge-spin Chern number $\mathcal C_{\rm ch}^{\rm sp} = \sum_{ss'} \mathcal C_{ss'} m_{s'}$, given that the relation $\mathcal C_{\rm sp}^{\rm ch} = \mathcal C^{\rm sp}_{\rm ch}$ holds.
Finally, the spin-spin Chern number $\mathcal C_{\rm sp}^{\rm sp} = \sum_{ss'} m_s \mathcal C_{ss'} m_{s'}$ is the third invariant.
Here $m_s$ are spin (generalized helicity) quantum numbers where $m_+ = 1$ and $m_- = -1$.

Given that our $(1\,+\,1)$-dimensional system involves a real
dimension $x$ and a synthetic dimension $\tilde y$, it is useful to relate our results to standard SU(2) (spin-1/2) condensed-matter physics of electrons and holes in two dimensions, that is, in two real spatial dimensions $(x,y)$.
For 2D systems, the charge or spin (generalized helicity) current is $J_y^{\lambda}$ in the $x$ direction, with $\lambda = \{{\rm ch}, {\rm sp}\}$.
The off-diagonal component of the conductivity tensor $\bar{\sigma}^{\lambda \tau}_{yx}$ is connected to $J_y^{\lambda}$ via the relation $J_{y}^{\lambda} = \bar{\sigma}^{\lambda \tau}_{yx} E_x^\tau$, where $E_x^\tau$ is a generalized electric field with $\tau = \{{\rm ch}, {\rm sp}\}$ \cite{haldane-2006}.
The field $E_{x}^{\rm ch}$ represents the conventional electronic field and $E_x^{\rm sp}$ describes the gradient of a state-dependent Zeeman field or state-dependent chemical potential.
A simplification of our notation is obtained by dropping the $xy$ labels and defining the conductivity tensor as $\bar{\sigma}_{yx}^{\lambda \tau} \equiv \bar{\sigma}^{\lambda}_{\tau}$ and the conductance tensor as $\sigma_{yz}^{\lambda \tau} \equiv \sigma_{\tau}^{\lambda}$.
Considering spin-1/2 fermions with charge $e$, the charge-charge (charge-Hall) conductance is $\sigma_{\rm ch}^{\rm ch} = (e^2/h)\mathcal C_{\rm ch}^{\rm ch}$, the spin-charge (spin-Hall) conductance is $\sigma^{\rm ch}_{\rm sp} = (e^2/h)(\hbar/e)\mathcal C_{\rm sp}^{\rm ch} = (e/2\pi)\mathcal C_{\rm sp}^{\rm sp}$.
However, our fermions are neutral and their spins represent two internal states of the atoms, which exist in a $(1\ +\ 1)$-dimensional space.
Thus, measuring charge-charge (charge-Hall), spincharge (spin-Hall), and spin-spin Chern numbers can be performed via standard proposed schemes \cite{satija-2011, cooper-2012, goldman-2013} or via actual experiments of Chern numbers for atomic systems with one and two internal states \cite{esslinger-2014, goldman-2015}.

In \ref{ch3table1} the charge-charge Chern number $\mathcal C_{\rm ch}^{\rm ch}$ is represented by $\mathcal C$, the spin-charge $\mathcal C^{\rm ch}_{\rm sp}$ is represented by $\chi$, and the topological invariant $\mathcal C^{\rm sp}_{\rm sp}$ is not shown, because it is equivalent to $\mathcal C^{\rm ch}_{\rm ch}$ for spin-1/2 systems \cite{sa-de-melo-2021}.
Notice that $\mathcal C^{\rm sp}_{\rm sp}$ does not add any additional information about the topological nature of insulating phases for SU(2) systems; it is sufficient to stop the topological classification at the spin-charge (spin-Hall) level, such as the $Z_2$ classification used in the case of quantum spin-Hall phases of graphenelike structures \cite{mele-2005, sheng-2011}.

The connection between our neutral $(1\,+\,1)$-dimensional
system and a charged 2D system is quite remarkable and
shows the intimate relationship between topological insulators in bichromatic lattices in one dimension and quantum
Hall systems in two dimensions via the mapping outlined below Eq.~[\ref{eq3.27}].

\section{Comparison with other work}
\label{sec3.7}

The effects of spin-orbit coupling in producing mobility edges for two-dimensional tight-binding fermions in a square lattice have been investigated \cite{tobe-2008}.
In that work the spin orbit is of the Rashba-type, the system has an external magnetic field perpendicular to the plane of the lattice, and the localization properties are studied as a function of the Rashba coupling $\lambda_R$ and the hopping anisotropy $\lambda_H = t_y/t_x$.
For fixed $\lambda_R$ and changing anisotropy $\lambda_H$, their system exhibits four phases:
For small $\lambda_H$ all states are localized (phase I), for
intermediate $\lambda_H$ states can be localized or extended (phases II and III), and for large $\lambda_H$ all states are extended.
In their phase diagram there is no periodicity in $\lambda_R$, there is no external Zeeman (Rabi) field, and there is no discussion about topological aspects associated with edge states and disorder.
Furthermore, the effects of spin-orbit coupling in producing mobility edges for two-dimensional Fermi systems have been investigated in continuum models when speckle potentials are present \cite{orso-2017}.
In that work the mobility edge was studied as a function the Rashba $\lambda_R$ and Dresselhaus $\lambda_D$ spin-orbit coupling for fixed speckle disorder amplitude $V_0$.
In that system, there is no periodicity in $\lambda_R$ or $\lambda_D$, there is no external Zeeman (Rabi) field, and there is no discussion associated with edge states and disorder.

In contrast, our work discusses quasiperiodic potentials in one spatial dimension in the presence of spin-orbit coupling and Rabi fields.
We find that mobility regions emerge due to the breaking of duality introduced by the simultaneous presence of spin-orbit coupling and Rabi (Zeeman) fields.
We show that the existence of spin-orbit coupling in one dimension, with zero Rabi (Zeeman) field, is not sufficient to produce mobility regions, because the spin-gauge symmetry removes the spin-orbit coupling from the problem.
However, the presence of a simultaneous Rabi field and spin-orbit coupling leads to mobility regions and to three-types of phases:
(a) All states are localized, (b) some states are localized and
others are extended, and (c) all states are localized.
We compute the inverse participation ratio to obtain phase diagrams showing regions of localized and extended states in the plane of the Aubry-Andr{\'e} parameter $\Delta/J$ versus spin-orbit coupling $k_T a$ or versus Rabi (Zeeman) field $h_x/J$.
We show that the phase diagrams are $\pi$ periodic in $k_T a$ and that localization can occur below the Aubry-Andr{\'e} threshold $(\Delta / J)_c = 2$.
Furthermore, a fundamental difference between our work and that of \cite{tobe-2008,orso-2017} is that their systems do not have edges and therefore there are no edge states.
Those authors were concerned only with bulk properties.
An analysis of edge states and of a generalized bulk-edge correspondence (via the Chern matrix) shows that the bulk-localized phases can be topological, possessing not only charge-charge but also spin-charge Chern numbers in analogy to two-dimensional bulk band-gap insulators.

\section{Conclusions}
\label{sec3.8}

We have found that the Aubry-Andr{\'e} model exhibits mobility regions when self-duality is broken, in contrast to the absence of mobility edges when global self-duality is preserved.
We discussed an explicit realization of dualitybreaking terms using spin-orbit coupling and Rabi fields in atomic wires with fermions.
We used parameters compatible with ${}^{40}{\rm K}$ and studied transitions from extended to localized phases as a function of disorder, spin-orbit coupling, Rabi fields, and filling factor.
We found three classes of nontrivial phases:
conventional topological insulator, conventional topological Aubry-Andr{\'e} insulator, and unconventional (hybrid) topological Aubry-Andr{\'e} insulator.
To obtain the topological invariants of these phases, we extended a one-dimensional family of Aubry-Andr{\'e} systems into a synthetic dimension and mapped it into the two-dimensional Harper-Hofstadter model used to describe quantum Hall insulators.
Our work paves the way for the experimental realization of the intriguing unconventional (hybrid) topological Aubry-Andr{\'e} insulators, in which edge states migrate from a localized to an extended bulk band and vice versa.

\bibliography{References}% Produces the bibliography via BibTeX.

\end{document}